\documentclass[acmtog]{acmart}
\acmSubmissionID{825}

\usepackage{booktabs} 

\usepackage{cancel}
\usepackage{enumitem}
\usepackage{amsmath}
\usepackage{relsize}
\usepackage{makecell}
\usepackage[symbol]{footmisc}
\usepackage{bm}
\usepackage{listings}
\usepackage{tcolorbox}
\usepackage{xcolor}
\usepackage{overpic,pbox,tikz}
\usepackage{multirow,colortbl}
\usepackage{siunitx}
\usepackage[utf8x]{inputenc}
\tcbuselibrary{listings, skins}
\usepackage[export]{adjustbox}
\usepackage[symbol]{footmisc}

\usepackage{pifont} 

\definecolor{myblue}{HTML}{4285f4}
\definecolor{myred}{HTML}{EA4335}
\definecolor{teaser_blue}{HTML}{00B0F0}
\definecolor{teaser_green}{HTML}{92D050}

\newcommand{\XSolidBrush}{\ding{55}}
\newcommand{\Checkmark}{\ding{51}} 
\newcommand{\xmark}{\texttt{\textcolor{myred}{\textbf{\XSolidBrush}}}}
\newcommand{\cmark}{\texttt{\textcolor{myblue}{\textbf{\Checkmark}}}}
\newcommand{\xtext}[1]{\texttt{\textcolor{myred}{\textbf{#1}}}}
\newcommand{\ctext}[1]{\texttt{\textcolor{myblue}{\textbf{#1}}}}

\definecolor{revision_blue}{rgb}{0.2,0.2,0.7}

\definecolor{darkgreen}{rgb}{0.0, 0.5, 0.0}

\definecolor{lightgray}{RGB}{240,240,240}
\definecolor{darkgray}{RGB}{64,64,64}
\definecolor{darkgray}{RGB}{64,64,64}
\definecolor{solarized_base03}{RGB}{  0,      43,     54}
\definecolor{solarized_base02}{RGB}{  7,      54,     66}
\definecolor{solarized_base01}{RGB}{  88,     110,    117}
\definecolor{solarized_base00}{RGB}{  101,    123,    131}
\definecolor{solarized_base0}{RGB}{   131,    148,    150}
\definecolor{solarized_base1}{RGB}{   147,    161,    161}
\definecolor{solarized_base2}{RGB}{   238,    232,    213}
\definecolor{solarized_base3}{RGB}{   253,    246,    227}
\definecolor{solarized_yellow}{RGB}{  181,    137,    0}
\definecolor{solarized_orange}{RGB}{  203,    75,     22}
\definecolor{solarized_red}{RGB}{     220,    50,     47}
\definecolor{solarized_magenta}{RGB}{ 211,    54,     130}
\definecolor{solarized_violet}{RGB}{  108,    113,    196}
\definecolor{solarized_blue}{RGB}{    38,     139,    210}
\definecolor{solarized_cyan}{RGB}{    42,     161,    152}
\definecolor{solarized_green}{RGB}{   133,    153,    0}

\lstdefinestyle{pseudocodestyle}{
    language=Python,
    basicstyle=\ttfamily\footnotesize\color{solarized_base02},    
    commentstyle=\color{solarized_cyan},
    keywordstyle=\bfseries\color{solarized_base03},
    numberstyle=\tiny\color{solarized_base03},
    stringstyle=\color{solarized_green},    
    tabsize=2,
    breaklines=true,
    numbers=left,       
    captionpos=t,  
    escapeinside={(*@}{@*)},
}

\AtBeginDocument{%
\newtcblisting[blend into=listings]{pseudocode}[1][]{%
  arc=2pt,
  boxrule=1pt,
  colframe=solarized_base00,
  enhanced,
  listing only,
  overlay={
    \begin{tcbclipinterior}
        \fill[solarized_base00!25](frame.south west)
        rectangle ([xshift=10pt]frame.north west);
    \end{tcbclipinterior}},
  listing remove caption=false,
  listing options={numbers=left, numberstyle=\tiny,   
    basicstyle=\small\ttfamily, 
    xleftmargin=2pt, 
    aboveskip=\smallskipamount, 
    belowskip=\smallskipamount,
    columns=fullflexible,
    style=pseudocodestyle
  },
  coltitle=black,
  attach boxed title to bottom center={yshift=-0pt},
  boxed title style={enhanced jigsaw, colback=white, sharp corners, boxrule=0pt},
  #1
}
}

\usepackage[ruled]{algorithm2e} 

\SetAlFnt{\small}
\SetAlCapFnt{\small}
\SetAlCapNameFnt{\small}
\SetAlCapHSkip{0pt}

\acmJournal{TOG}
\acmYear{2025} 
\acmVolume{44} 
\acmNumber{4} 
\acmArticle{} 
\acmMonth{8}

\setcopyright{acmlicensed}

\acmDOI{10.1145/3730892}

\begin{document}
\title{TransparentGS: Fast Inverse Rendering of Transparent Objects with Gaussians}

\author{Letian Huang}
\orcid{0009-0003-1454-7824}
\affiliation{%
 \institution{State Key Lab for Novel Software Technology, Nanjing
University}
 \city{Nanjing}
 \country{China}}
\email{lthuang@smail.nju.edu.cn}

\author{Dongwei Ye}
\orcid{0009-0004-8637-4384}
\email{dongweiye@smail.nju.edu.cn}
\affiliation{%
 \institution{State Key Lab for Novel Software Technology, Nanjing
University}
 \city{Nanjing}
 \country{China}}
 
\author{Jialin Dan}
\orcid{0009-0007-2228-4648}
\email{danjialin@smail.nju.edu.cn}
\author{Chengzhi Tao}
\orcid{0000-0002-0736-7951}
\email{tcz_tao@smail.nju.edu.cn}
\affiliation{%
 \institution{State Key Lab for Novel Software Technology, Nanjing
University}
 \city{Nanjing}
 \country{China}}

\author{Huiwen Liu}
\orcid{0009-0005-6423-4812}
\affiliation{%
 \institution{TMCC, College of Computer Science, Nankai University}
 \city{Tianjin}
 \country{China}}
\email{huiwenliu@mail.nankai.edu.cn}

\author{Kun Zhou}
\orcid{0000-0003-4243-6112}
\affiliation{%
 \institution{State Key Lab of CAD \& CG, Zhejiang University}
 \city{Hangzhou}
 \country{China}
 }
 \affiliation{%
 \institution{Institute of Hangzhou Holographic
Intelligent Technology}
 \city{Hangzhou}
 \country{China}
 }
\email{kunzhou@zju.edu.cn}

\author{Bo Ren}
\orcid{0000-0001-8179-9122}
\affiliation{%
 \institution{TMCC, College of Computer Science, Nankai University}
 \city{Tianjin}
 \country{China}}
\email{rb@nankai.edu.cn}

\author{Yuanqi Li}
\orcid{0000-0003-4100-7471}
\email{yuanqili@nju.edu.cn}
\author{Yanwen Guo}
\orcid{0000-0002-7605-5206}
\email{ywguo@nju.edu.cn}
\affiliation{%
 \institution{State Key Lab for Novel Software Technology, Nanjing
University}
 \city{Nanjing}
 \country{China}}

\author{Jie Guo}
\orcid{0000-0002-4176-7617}
\authornote{Corresponding authors.}
\affiliation{%
 \institution{State Key Lab for Novel Software Technology, Nanjing
University}
 \city{Nanjing}
 \country{China}}
\email{guojie@nju.edu.cn}


%
%


%
%


\begin{teaserfigure}
	\newlength{\lenTeaser}
	\setlength{\lenTeaser}{\linewidth}
	\addtolength{\tabcolsep}{-5pt}
    \center
	\begin{overpic}[width=\lenTeaser]{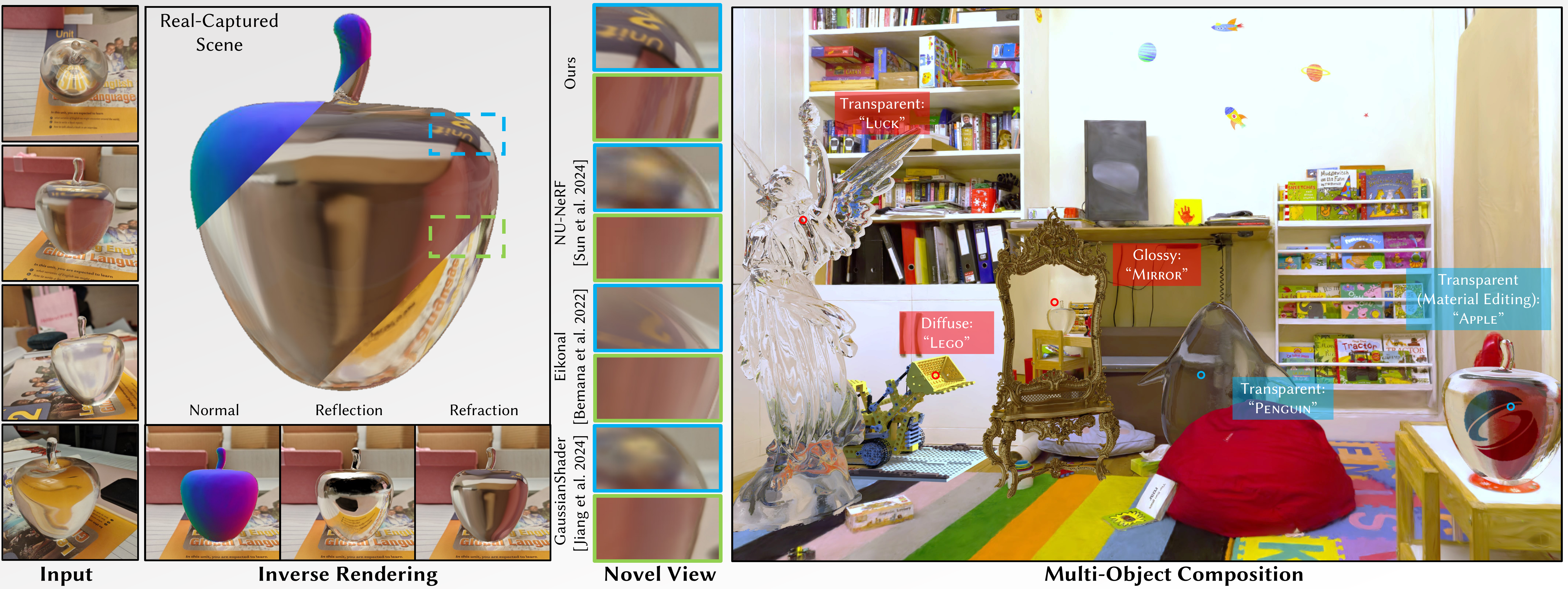}
	\end{overpic}
	\caption{\label{fig:teaser}
        TransparentGS is a novel inverse rendering pipeline based on 3D-GS, specifically designed for reconstructing transparent objects under various light and viewing conditions. It delivers high-quality reconstruction results within \textbf{1 hour}, enabling real-time novel view synthesis and convenient re-rendering of scenes with challenging transparent objects. Compared to state-of-the-art methods, it achieves both high-fidelity refraction (\texttt{\textcolor{teaser_blue}{\textbf{blue}}} box) and reflection (\texttt{\textcolor{teaser_green}{\textbf{green}}} box) with nearby contents (indirect light), thereby paving the way for complex secondary ray effects in the scene composited of multiple objects.
		}
\end{teaserfigure}

\begin{abstract}
The emergence of neural and Gaussian-based radiance field methods has led to considerable advancements in novel view synthesis and 3D object reconstruction. Nonetheless, specular reflection and refraction continue to pose significant challenges due to the instability and incorrect overfitting of radiance fields to high-frequency light variations. Currently, even 3D Gaussian Splatting (3D-GS), as a powerful and efficient tool, falls short in recovering transparent objects with nearby contents due to the existence of apparent secondary ray effects. To address this issue, we propose TransparentGS, a fast inverse rendering pipeline for transparent objects based on 3D-GS. The main contributions are three-fold. Firstly, an efficient representation of transparent objects, transparent Gaussian primitives, is designed to enable specular refraction through a deferred refraction strategy. Secondly, we leverage Gaussian light field probes (GaussProbe) to encode both ambient light and nearby contents in a unified framework. Thirdly, a depth-based iterative probes query (IterQuery) algorithm is proposed to reduce the parallax errors in our probe-based framework. Experiments demonstrate the speed and accuracy of our
approach in recovering transparent objects from complex environments, as well as several applications in computer graphics and vision.
\end{abstract}

\maketitle

\begin{table}[t]
    \setlength{\tabcolsep}{2pt} 
\caption{\label{tab:teaser}\textbf{Comparison of transparent object reconstruction
methods} in terms of training time \textbf{(A)}, rendering time \textbf{(B)}, and the capability of supporting ambient light \textbf{(C)}, nearby contents (indirect light) \textbf{(D)}, high-frequency refraction details \textbf{(E)}, accurate reflection-refraction decoupling \textbf{(F)}, colored refraction \textbf{(G)}, and re-rendering (e.g., relighting or material editing~\cite{khan2006image}) \textbf{(H)}.
}

\resizebox{\linewidth}{!}{%
\begin{tabular}{c|cccccccc}
\begin{tabular}[c]{@{}c@{}}\textbf{Methods}\end{tabular} & 
\begin{tabular}[c]{@{}c@{}}{\textbf{A}}\end{tabular} &
\begin{tabular}[c]{@{}c@{}}{\textbf{B}}\end{tabular} 
&\begin{tabular}[c]{@{}c@{}}{\textbf{C}}\end{tabular} & \begin{tabular}[c]{@{}c@{}}{\textbf{D}}\end{tabular} & \begin{tabular}[c]{@{}c@{}}{\textbf{E}}\end{tabular} & \begin{tabular}[c]{@{}c@{}}{\textbf{F}}\end{tabular} &\begin{tabular}[c]{@{}c@{}}{\textbf{G}}\end{tabular} & \begin{tabular}[c]{@{}c@{}}{\textbf{H}}\end{tabular} \\
\toprule
Eikonal~\cite{bemana2022eikonal}   & \xtext{slow (>10h)} & \xtext{offline} & \cmark & \cmark & \xmark  & \xmark & \xmark & \xmark   \\
NEMTO~\cite{wang2023nemto}  & \xtext{slow (>10h)}  & \xtext{offline}   & \cmark   & \xmark   & \cmark    & \cmark &\xmark & \cmark  \\
\citet{gao2023transparent}   & \xtext{slow (>10h)}  & \xtext{offline}  & \xmark   & \cmark   & \cmark  & \xmark & \xmark  & \cmark \\
NU-NeRF~\cite{sun2024nu} & \xtext{slow (>7h)}  & \xtext{offline}  & \cmark   & \cmark   & \xmark  & \cmark & \xmark & \cmark  \\
\cmidrule(lr){1-9}
\textbf{TransparentGS (Ours)}   & \ctext{fast (<1h)}    & \ctext{real-time}   & \cmark  & \cmark  & \cmark      & \cmark  & \cmark
  & \cmark \\
\bottomrule
\end{tabular}
}

\end{table}

\section{Introduction}

Reconstructing 3D scenes from multiple 2D images of different views and synthesizing novel views has been a long-standing task in computer graphics and vision. With the advent of deep learning, the trend in solving this task has been spearheaded by Neural Radiance Fields (NeRFs)~\cite{mildenhall2020nerf} and its variants~\cite{barron2022mip, muller2022instant}, which achieve photorealistic visual quality by using
volume rendering with implicit fields. Recently, as the demand for real-time rendering continues to increase, 3D Gaussian Splatting (3D-GS)~\cite{kerbl20233d} offers a more efficient explicit representation that
can achieve real-time rendering by modeling radiance fields as 3D Gaussian primitives. 

Unfortunately, until now it remains challenging to reconstruct transparent objects under arbitrary light conditions and enable real-time novel view synthesis. The challenge arises from the frequent appearance variations of transparent objects across different viewpoints, due to the intricate interplay of reflection and refraction as light traverses the material. Consequently, both MLPs with the directional encodings in NeRF and the spherical harmonic functions (SH) in 3D-GS face difficulties in modeling specular reflection and refraction accurately.

Currently, the most effective reconstruction methods for transparent objects are based on implicit neural representations~\cite{li2020through, bemana2022eikonal, gao2023transparent, sun2024nu}. However, these method are notorious for their significant computational overhead during training, and also do not facilitate real-time rendering, as explained and compared in Tab.~\ref{tab:teaser}. 3D-GS-based methods offer high efficiency in 3D reconstruction and novel view synthesis~\cite{wu2024recent}, but existing methods have their own limitations. In particular, these methods~\cite{jiang2024gaussianshader, ye20243defer}, originally designed for reflective scenes, cannot be easily applied to transparent objects. The difficulty is exacerbated when the transparent object is surrounded by nearby objects. Due to the rasterization-based formulation, 3D-GS is constrained to ideal pinhole cameras and lacks support for secondary ray effects such as refraction and inter-reflection~\cite{letian2024op43dgs, moenne20243dtracing}.

In this paper, we propose \emph{TransparentGS}, a novel inverse rendering framework that builds upon 3D-GS to efficiently manage transparent objects in a variety of light and viewing conditions, supporting both refraction and inter-reflection with nearby contents. In this framework, transparent objects are modeled by \emph{transparent Gaussian primitives}, which explicitly encode both geometric and material properties within 3D Gaussians. These new primitives are convenient to be integrated into a dedicated \emph{physically-based deferred rendering pipeline}, enabling the consolidation of mesh and GS for secondary ray effects. To handle both ambient light and nearby contents in the scene, we introduce \emph{Gaussian light field probes (GaussProbe)} to capture the local light field of each transparent object by placing a sparse set of caches around it. These probes can be baked and updated fastly based on an optimal projection strategy~\cite{letian2024op43dgs}. To address the parallax issue inherent to the probes and enhance the details of refraction/inter-reflection, we design a \emph{depth-based iterative probes query algorithm (IterQuery)} which achieves plausible results only after a few iterations. 

Compared to previous work, our new framework has several advantages as listed in Tab.~\ref{tab:teaser}. In particular, we achieve the first fast inverse rendering with secondary ray effects of transparent objects \emph{within one hour}, enabling real-time novel view synthesis of transparent scenes and complex secondary ray effects in the interplay, as illustrated in Fig.~\ref{fig:teaser}. 

In summary, the contributions of this work are as follows:
\begin{itemize}
    \item We design an efficient representation of transparent objects, transparent Gaussian primitives, which enables specular refraction through a deferred refraction strategy.
    \item We leverage Gaussian light field probes (GaussProbe) to encode both ambient light and nearby contents within a unified framework.
    \item We propose a depth-based iterative probe query algorithm (IterQuery) to mitigate the parallax errors in our probe-based framework.
\end{itemize}

\section{Related Work} 

In this section, we briefly review the reconstruction and novel view synthesis methods for transparent objects, and also discuss various light representations in inverse rendering.

\paragraph{Transparent object reconstruction}
To reconstruct transparent objects, a few early methods~\cite{stets2017scene, huynh2010shape, wetzstein2011refractive, shao2024polarimetric, ihrke2010transparent, qian20163d, kutulakos2008theory} employ dedicated hardware setups like polarization cameras or require additional information beyond a simple set of images. Subsequent works relax the constraints but still rely on unique patterns to infer the correspondence between the camera ray and the background~\cite{li2023neto, lyu2020differentiable, xu2022hybrid, wu2018full} or require manual annotations~\cite{li2020through, xu2022hybrid}. Recently, with the emergence of neural radiance fields~\cite{mildenhall2020nerf} in novel view synthesis tasks, numerous studies leverage neural implicit representations to achieve transparent object reconstruction~\cite{zhan2023nerfrac, tong2023seeing, gao2023transparent, sun2024nu}. \citet{bemana2022eikonal} propose novel view synthesis by learning an IoR field alongside radiance and density, addressing the bent light path in the field without modeling the surface geometry of objects. The eikonal equation
deals with refraction and total internal reflection but not separation into partial reflection and refraction, which poses challenges for re-rendering. NEMTO~\cite{wang2023nemto} employs a neural network to predict refracted directions and performs physically-based shading under the assumption of ambient light. \citet{gao2023transparent} propose a two-stage method to predict multi-view silhouettes and reconstruct the shape of a refractive object on an opaque plane, which serves as both a geometry and appearance prior. Both methods~\cite{wang2023nemto,gao2023transparent} are limited by their light representations and cannot handle realistic scenes that simultaneously include nearby objects and light. NU-NeRF~\cite{sun2024nu} employs a two-stage reconstruction strategy to handle nested transparent objects. However, it utilizes a network to predict refracted light, making it without refraction details. Moreover, these neural rendering methods are very time-consuming due to the intensive network inference and sampling required. Currently, all the aforementioned neural methods require far more time than our method to reconstruct transparent objects.

\paragraph{Light representation} When reconstructing specular or transparent objects, it is essential to account for their surrounding environment to accurately analyze the reflected and refracted light paths. Previous methods for novel view synthesis of transparent objects~\cite{wang2023nemto} and inverse rendering of opaque objects~\cite{jiang2024gaussianshader,lai2024glossygs,liang2023envidr} typically assume that all light comes 
from an infinitely faraway environment and approximate it using an environment maps, which neglects secondary ray effects. To model indirect light, TensoIR~\cite{jin2023tensoir}, based on a tensor-factorized representation~\cite{chen2022tensorf}, explicitly computes ray integrals to ensure accurate visibility and indirect light in radiance field rendering, which is highly time-consuming. NeILF++~\cite{zhang2023neilf++} proposes to marry an incident light field and an outgoing radiance field through physically-based rendering, which enables the handling of specular surfaces and inter-reflection, but often results in overly smooth outcomes and requires extensive training time. \citet{gao2023transparent} assume all refracted light comes from a perfect textured plane, overlooking distant ambient light. Following NeRO~\cite{liu2023nero}, NU-NeRF~\cite{sun2024nu} employs two MLPs with Integrated Directional Encoding (IDE)~\cite{verbin2022ref} to model the ambient light and indirect inter-reflected light from nearby objects, respectively. It further utilizes an MLP that simultaneously inputs position and view direction to predict refracted light, leading to over-blurred results. NeLF-Pro~\cite{you2024nelf} models a scene using light field probes in the context of NeRF~\cite{mildenhall2020nerf}. It primarily models outgoing radiance rather than incident light, which makes it challenging to directly query the incident light during the inverse rendering process. TraM-NeRF~\cite{holland2024tram} traces reflected rays and introduces a radiance estimator that combines volume and reflected radiance integration to model inter-reflection. However, it cannot handle refractive objects and does not support real-time rendering. \citet{liang2024gs} and \citet{gao2025relightable} employ spherical harmonics (SH) to model indirect light for opaque objects, which is not suitable for representing the high-frequency details of specular refraction. \citet{zhou2024unified} propose a novel and insightful Gaussian-based offline path tracer, which, however, does not support fast inverse rendering of transparent objects. Our proposed Gaussian light field probes, on the other hand, can model indirect light with high fidelity and performance for specular refraction and inter-reflection.

\section{Method}

\begin{figure*}[h]
\centering
\includegraphics[width=1\textwidth]{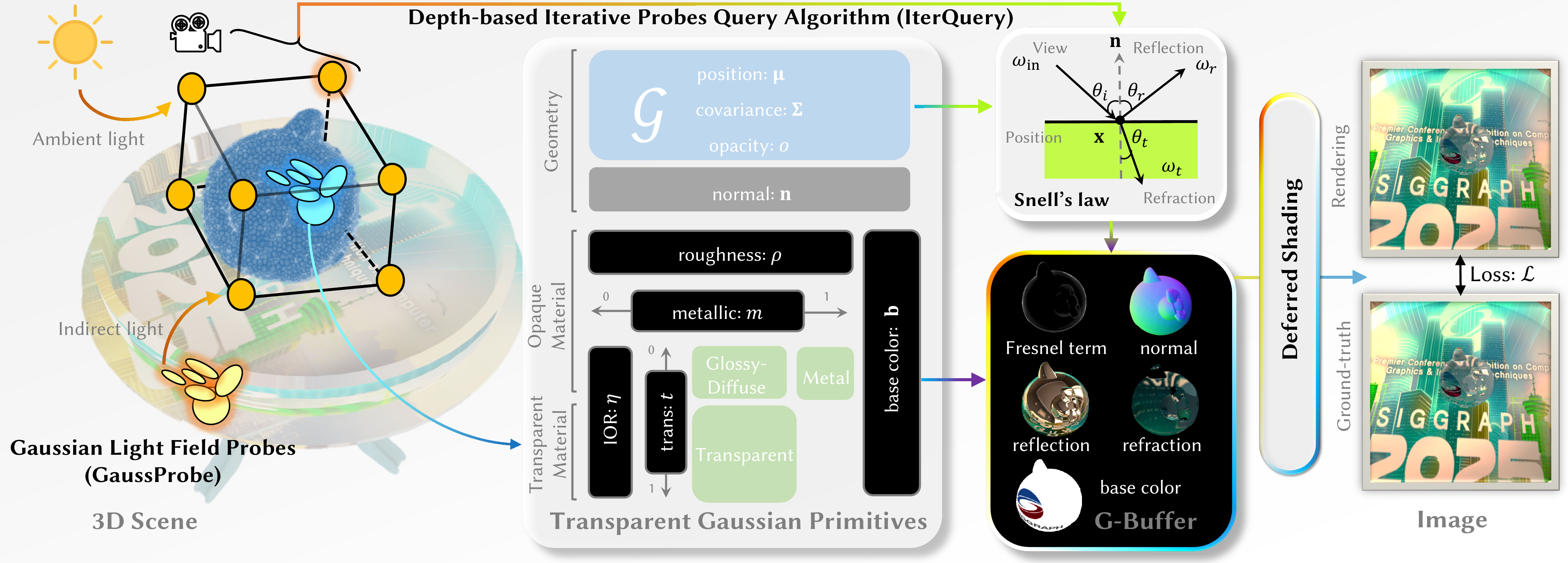}
\caption{\textbf{The overview of our TransparentGS pipeline.} Each 3D scene is firstly separated into transparent objects and opaque environment using SAM2~\cite{ravi2024sam} guided by GroundingDINO~\cite{liu2025grounding}. For transparent objects, we propose transparent Gaussian primitives, which explicitly encode both geometric and
material properties within 3D Gaussians. And the properties are rasterized into maps for subsequent deferred shading. For the opaque environment, we recover it with the original 3D-GS, and bake it into GaussProbe surrounding the transparent object. The GaussProbe are then queried through our IterQuery algorithm to compute reflection and refraction.} 
\label{fig:overview}
\end{figure*}

In this section, we present our method, TransparentGS, tailored for the fast inverse rendering of transparent objects. As depicted in Fig.~\ref{fig:overview}, our TransparentGS consists of two parts. The first is the representation of transparent objects, transparent Gaussian primitives, which encompasses shape attributes following 3D Gaussian Splatting (3D-GS)~\cite{kerbl20233d}. To enhance the representation capability of refraction, we extend the primitives with additional attributes, including normals and parameters of the Bidirectional Scattering Distribution
Function (BSDF)~\cite{bartell1981theory}. To render detailed specular refraction with the primitives, we employ a deferred refraction method. The second is our proposed Gaussian light field probes (GaussProbe), which not only represent ambient light similar to environment maps but also indirect light from nearby contents. To eliminate parallax issues, we develop a depth-based iterative probes query (IterQuery) algorithm dedicated to Gaussian light field probes.

\subsection{Preliminaries}

\paragraph{3D Gaussian Splatting} 3D Gaussian Splatting (3D-GS)~\cite{kerbl20233d} constructs a scene using 3D Gaussian primitives $\mathcal{G}\left(\cdot\right)$, each characterized by position $\boldsymbol{\mu}$, a covariance matrix $\boldsymbol{\Sigma}$, opacity $o$, and spherical harmonics coefficients $\text{SH}\left(\cdot\right)$ representing view-dependent appearance. To render an image, the 3D Gaussian is transformed into the camera space with world-to-camera transform matrix $\mathbf{W}$ and projected to 2D image space through a local affine approximation Jacobian matrix $\mathbf{J}$~\cite{zwicker2002ewa}:
\begin{gather}
\boldsymbol{\Sigma}_{\text{2D}}=\mathbf{J}\mathbf{W}\boldsymbol{\Sigma}\mathbf{W}^{\top}\mathbf{J}^{\top}
\end{gather} where $\boldsymbol{\Sigma}_{\text{2D}}$ is the covariance matrix of the projected Gaussian $\mathcal{G}^{\text{2D}}$.
Next, 3D-GS~\cite{kerbl20233d} employs volumetric \(\alpha\)-blending with weights $\alpha=o\mathcal{G}^{\text{2D}}\left(\cdot\right)$ to integrate alpha-weighted appearance. The formulation is similar to NeRF~\cite{mildenhall2020nerf}:
\begin{equation}
    \mathcal{C}=\sum_{i=1}^{N}T_i\alpha_i\text{SH}_i\left(\mathbf{v}\right)
\label{eq:alpha-blending}
\end{equation} where $N$ is the number of primitives within a pixel, and $i$ is the index of the primitives. $\mathbf{v}$ denotes the view direction, and $T_i=\prod_{j=1}^{i-1}\left(1-\alpha_j\right)$ is the accumulated transmittance.

\paragraph{Rendering equation} To achieve physically photorealistic rendering of surface geometry, we follow the rendering equation~\cite{kajiya1986rendering} accounting for both reflection and transmission, and employ the bidirectional scattering distribution function (BSDF):
\begin{equation}
L=\int_{\Omega}L_{\text{in}}\left(\mathbf{x},\omega_{\text{in}}\right)f\left(\mathbf{x},\omega,\omega_{\text{in}}\right)\left|\omega_{\text{in}}\cdot\mathbf{n}\right|~\mathrm{d}\omega_{\text{in}}.
\label{eq:rendering_eq}
\end{equation} \(L\) is the shading result, \(L_{\text{in}}\) is the incident radiance, \(\mathbf{n}\) is the surface normal, and \(\omega, \omega_{\text{in}}\) are the outgoing and incident directions. \(f\) denotes the BSDF at surface point \(\mathbf{x}\) which can be further subdivided as the bidirectional reflectance distribution function (BRDF) and the bidirectional transmittance distribution function (BTDF).

\subsection{Transparent Gaussian Primitives}
\label{sec:transparent_primitives}

The spherical harmonics in 3D-GS~\cite{kerbl20233d} struggle to accurately capture the high-frequency variations of appearance in specular reflection and refraction, leading to either overfitting, which causes degradation in novel views, or highly blurred results. To address this, we propose a novel representation, transparent Gaussian primitives, that involves the physically-based rendering pipeline rather than computing spherical harmonics directly. 

\paragraph{Parameterization} As shown in Fig.~\ref{fig:overview}, we parameterize the surface geometry as normal \(\mathbf{n}\) and the surface material as roughness \(\rho\) and metallic \(m\). For transparent objects, it is essential to account for transmission and refraction at the surface. Thus, we utilize two additional material parameters: transparency $t$ and the index of refraction (IOR) $\eta$. The transparency $t$ is used to interpolate between the opaque and transparent material. Additionally, to preserve the high running performance, we retain the shape parameters of the original 3D-GS~\cite{kerbl20233d} for \(\alpha\)-blending and rasterization.

The opacity attribute plays a key role in 3D-GS~\cite{kerbl20233d} even for opaque objects, as 3D-GS relies on volumetric $\alpha$-blending. For transparent objects, merely using a lower opacity to represent transparency will fail to represent the surface, which in turn makes it hard to compute refraction and reflection. Therefore, we introduce an additional parameter $t$ to facilitate applications in computer graphics, such as material editing~\cite{khan2006image}.

\paragraph{Surface reflection and refraction}  In this paper, we explicitly represent surface reflection and refraction through two separate BSDFs:
\begin{equation}
    f=(1-t)f_r+tf_t
\end{equation} where $f_r$ denotes BRDF and $f_t$ denotes BTDF. For opaque materials with $t=0$, we follow the Cook-Torrance model~\shortcite{cook1981reflectance}. And for the reflective component of metals or transparent objects with $\rho=0$, we treat it as perfect specular reflection:
\begin{equation}
    f_r=F\frac{\delta\left(\omega - \omega_r\right)}{\left|\omega_{\text{in}}\cdot\mathbf{n}\right|},\quad\text{with $\omega_r=2\left(\omega_{\text{in}}\cdot\mathbf{n}\right)\mathbf{n}-\omega_{\text{in}}$}
\end{equation} where $\omega_r$ denotes the analytical reflected direction, $\delta$ denotes the Dirac delta function, and Fresnel term $F$ can be approximated using Schlick Approximation~\cite{schlick1994inexpensive}. Similarly, for the transmissive component of transparent objects with $\rho=0$, we consider only the perfect specular refraction:
\begin{equation}
    f_t=(1-F)\frac{\delta\left(\omega - \omega_t\right)}{\left|\omega_{\text{in}}\cdot\mathbf{n}\right|}
\label{eq:btdf}
\end{equation} where $\omega_t$ refers to the refracted direction. As depicted in Fig.~\ref{fig:overview}, the analytical refracted direction is obtained according to Snell's law.

\begin{figure}[!t]
  \centering
  \includegraphics[width=0.98\linewidth]{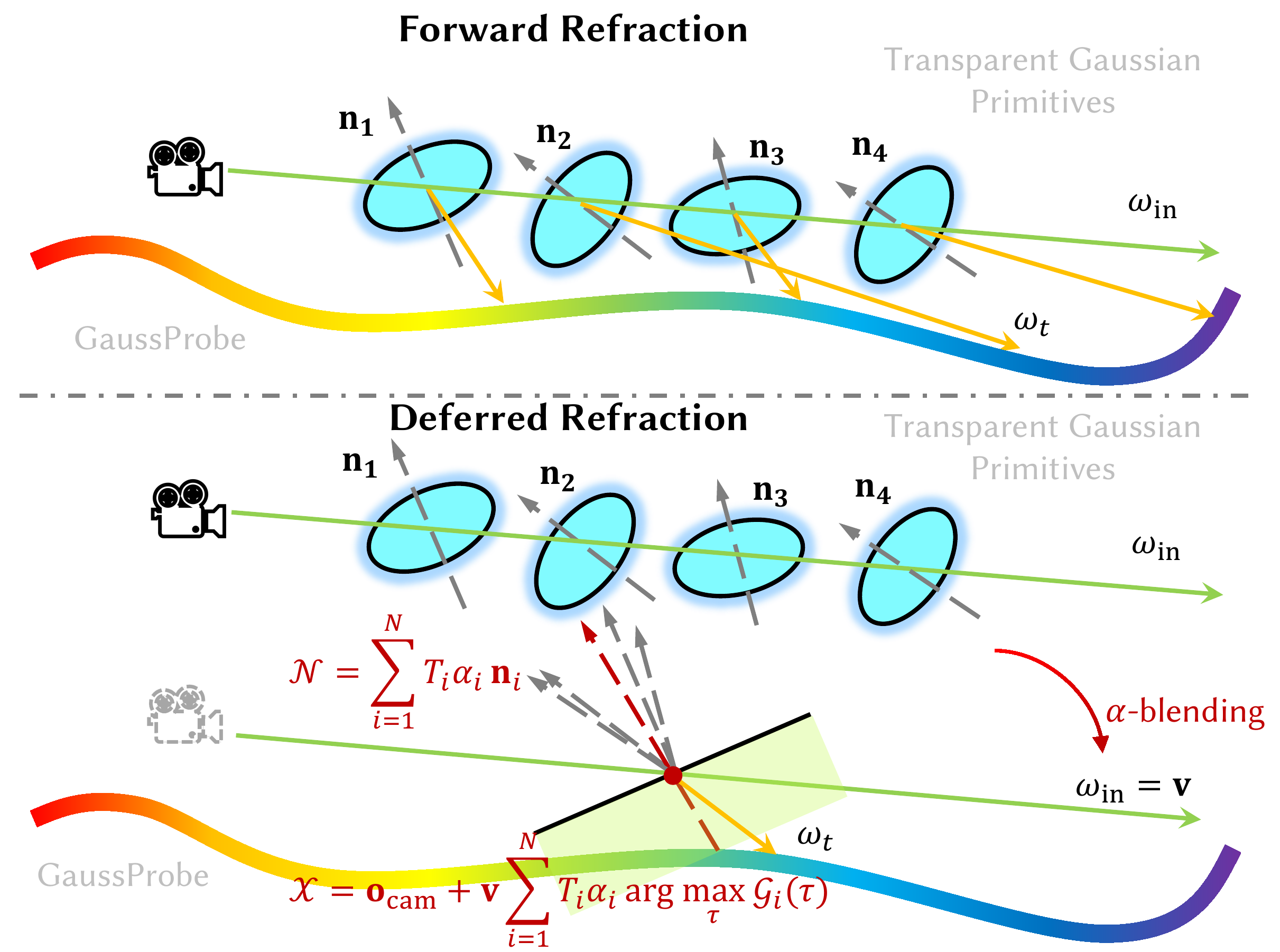} 
  \caption{\label{fig:deferred_shade}
    \textbf{Difference between forward and deferred refraction strategies.}  The gray arrows denote the normal attributes $\mathbf{n}_i$ of transparent Gaussian primitives (the blue ellipsoids). Deferred refraction integrates the alpha-weighted normal $\mathcal{N}$ and the alpha-weighted hitting point $\mathcal{X}$, and samples the GaussProbe with the single refracted ray $\omega_t$ (the orange arrow).
  }
\end{figure}

\paragraph{Deferred refraction} A critical aspect of employing 3D-GS for inverse rendering is the shading scheme. Generally, the shading strategy can be categorized into two types: forward shading~\cite{dihlmann2024subsurface,gao2025relightable} and deferred shading~\cite{deering1988triangle, ye20243defer, wu2024deferredgs}. “Forward” refers to performing shading first, followed by $\alpha$-blending, while “deferred” does the opposite. Prior studies~\cite{lai2024glossygs, ye20243defer} have highlighted the impact of operation order on specular reflection. Inspired by some work on deferred reflection~\cite{ye20243defer, wu2024deferredgs}, we explore the possibility of adopting deferred shading in handling specular refraction. Specifically, we derive the transmissive color using Eqn.~\eqref{eq:rendering_eq} and Eqn.~\eqref{eq:btdf}: 
\begin{equation}
\begin{split}
L_t&=\int_\Omega{L_{\text{in}}\left(\mathbf{x},\omega_{\text{in}}\right)\left(1-F\right)\delta\left(\omega-\omega_t\right)}~\mathrm{d}\omega_{\text{in}}\\
&=\left(1-F\right)L_{\text{in}}\left(\mathbf{x},\omega_t\right).
\label{eq:transmissive_light}
\end{split}
\end{equation}Since the $\alpha$-blending is a linear operation and Eqn.~\eqref{eq:transmissive_light} is not linear with respect to the normal $\mathbf{n}$, the operation order has a significant impact on the refraction. Forward refraction averages the shading results of Gaussians. Compared to forward refraction, deferred
shading is more adept at capturing transparent objects with specular refraction for its single sampling of the illumination, as shown in Fig.~\ref{fig:deferred_shade}. In our deferred refraction strategy (see Fig.~\ref{fig:overview}), we employ a point-based $\alpha$-blending approach, similar to Eqn~\eqref{eq:alpha-blending}, to aggregate all reflection- and refraction-related attributes carried by the primitives. For instance, the alpha-weighted normal is computed by:
\begin{equation}
    \mathcal{N} =\sum_{i=1}^{N}T_i\alpha_i\mathbf{n}_i.
\end{equation} Moreover, compared to reflection based on ambient light, we also need to integrate the ray's hitting points for refraction and reflection. However, simply integrating the centers of the primitives $\boldsymbol{\mu}$ neglects their anisotropy, leading to errors. We adopt a more accurate strategy to compute the hitting point. Considering the primary ray $\mathbf{o}_{\text{cam}}+\tau\mathbf{v}$ with the origin $\mathbf{o}_{\text{cam}}$ and the view direction $\mathbf{v}=\omega_{\text{in}}$, the Gaussian value on the ray can be represented as a function of the distance $\tau$:
\begin{equation}
    \mathcal{G}\left(\tau\right)=\exp\left({-\frac{1}{2}\left(\mathbf{o}_{\text{cam}}+\tau\mathbf{v}-\boldsymbol{\mu}\right)^{\top}\boldsymbol{\Sigma}^{-1}\left(\mathbf{o}_{\text{cam}}+\tau\mathbf{v}-\boldsymbol{\mu}\right)}\right).
\end{equation}
We analytically compute the maximum value of this function and aggregate the responses of the Gaussians along the ray as follows:
\begin{equation}
\begin{split}
\mathcal{X} = \mathbf{o}_{\text{cam}} +\mathbf{v} \sum_{i=1}^{N} T_i \alpha_i \arg\max_{\tau} \mathcal{G}_i\left(\tau\right),\\
\arg\max_{\tau} \mathcal{G}_i\left(\tau\right)=\frac{\left(\boldsymbol{\mu}_i-\mathbf{o}_{\text{cam}}\right)^{\top}\boldsymbol{\Sigma}^{-1}\mathbf{v}}{\mathbf{v}^{\top}\boldsymbol{\Sigma}^{-1}\mathbf{v}}
\label{eq:hitting_point}
\end{split}
\end{equation} where $\mathcal{X}$ is the alpha-weighted hitting point and $\mathcal{G}_i$ denotes the $i$-th primitive along the ray. Eqn.~\eqref{eq:hitting_point} considers the anisotropy of transparent Gaussian primitives, where different pixels correspond to different hitting points. The refracted rays $\omega_t$ and reflected rays $\omega_r$ are computed using these aggregated attributes stored in the geometry buffer (G-buffer) for subsequent queries.

\paragraph{Handling Colored Transparent Objects} Our method can be easily extended to handle absorption in colored transparent objects. We only need to modify Eqn~\eqref{eq:transmissive_light} by:
\begin{equation}
    L_t=(1-F)L_{\text{in}}\left(\mathbf{x},\omega_t\right)e^{-\sigma\left(\lambda\right) d}
    \label{eq:handling_color}
\end{equation} where \(\sigma(\cdot)\) is the absorption coefficient of the material, \(\lambda\) 
 is the wavelength of the light, and $d$ denotes the path length that the light travels through the material. We approximate the exponential term (transmittance) using the alpha-weighted base color $\sum_{i=1}^{N}{T_i\alpha_i\mathbf{b_i}}$, where $\mathbf{b}_i$ is an optimizable parameter encoded in each transparent Gaussian primitive. Thanks to this design, our method effectively decouples the refraction term from the transparent object's inherent color, which is a challenge that other methods struggle to handle.

\subsection{Gaussian Light Field Probes}
\label{sec:GaussProbe}
The choice of incident light representation significantly affects inverse rendering. Using an environment map with only two degrees of freedom to represent indirect light from nearby contents causes severe parallax issues. To overcome the challenge of incoherent rays that vanilla 3D-GS~\cite{kerbl20233d} fails to handle, we propose a novel light representation: Gaussian light field probes (GaussProbe). The important symbols in this section are demonstrated in Tab.~\ref{tab:notation}.

\begin{figure*}[!ht]
  \centering
  \includegraphics[width=1\linewidth]{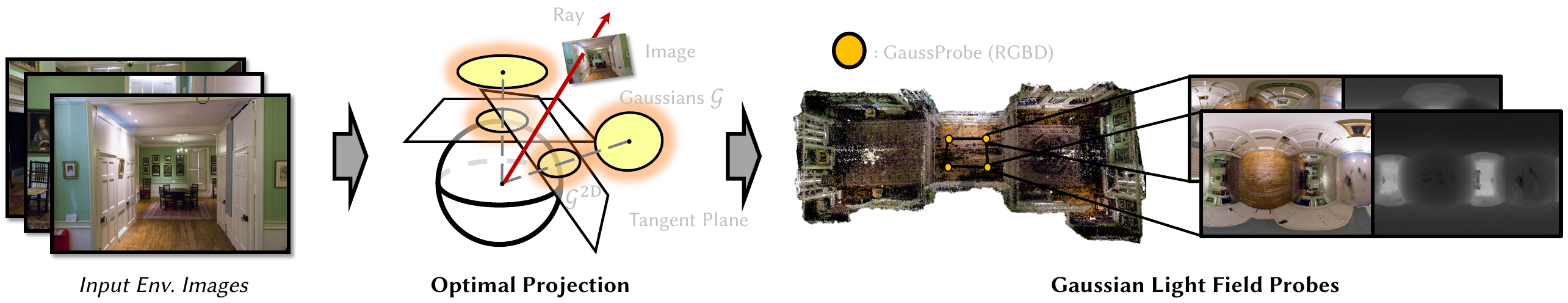} 
  \caption{\label{fig:probes_bake}
    \textbf{Illustration of our baking pipeline for Gaussian light field probes.} Given a set of environmental images with the transparent object removed, we can reconstruct the 3D scene using the original 3D-GS~\cite{kerbl20233d}. We voxelize the scene and place virtual cameras around the bounding box of the transparent object. For each virtual camera, we project the Gaussian primitives onto the tangent plane of the unit sphere, generating tangent-plane Gaussians. Finally, an $\alpha$-blending pass bakes the 360° panoramic color and depth maps at each point, which are subsequently stored in the voxels.
  }
\end{figure*}

\paragraph{Baking Gaussian light field probes} Before reconstructing transparent objects, we first bake the Gaussian light field probes. As shown in Fig.~\ref{fig:probes_bake}, we first reconstruct the environment via the vanilla 3D-GS~\cite{kerbl20233d}. We then voxelize the scene and place probes within the voxels surrounding the bounding box of the transparent object. For each probe, we follow the optimal projection~\cite{letian2024op43dgs} to render a panoramic image under the setting of a virtual camera. Specifically, we project the Gaussian primitives onto the corresponding tangent planes of the unit sphere with the projection function $\varphi$ instead of the image plane. The corresponding local affine approximation Jacobian matrix is modified by:
\begin{equation}
\mathbf{J}=\frac{1}{{(\mu_{x}^{2} + \mu_{y}^{2} + \mu_{z}^{2})^{\frac{3}{2}}}}\begin{bmatrix} {\mu_{y}^{2} + \mu_{z}^{2}} & -  {\mu_{x} \mu_{y}} & -  {\mu_{x} \mu_{z}}\\-  {\mu_{x} \mu_{y}} &  {\mu_{x}^{2} + \mu_{z}^{2}} & -  {\mu_{y} \mu_{z}}\\-  {\mu_{x} \mu_{z}} & -  {\mu_{y} \mu_{z}} &  {\mu_{x}^{2} + \mu_{y}^{2}}\end{bmatrix}
\end{equation} where $\boldsymbol{\mu}^{'}=\left[\begin{matrix}
    \mu_x,\mu_y,\mu_z
\end{matrix}\right]^{\top}$ denotes the position of 3D Gaussian in the camera space. Subsequently, for a pixel \((u,v)\) on the image, we cast a ray and compute the intersection with the tangent plane:
\begin{equation}
\mathbf{x}_{\text{2D}}=\varphi\left(\left[\begin{matrix}\sin{\left(\frac{\pi \left(- W + 2 u\right)}{W} \right)} \cos{\left(\frac{\pi \left(- H/2 + v\right)}{H} \right)}\\\sin{\left(\frac{\pi \left(- H/2 + v\right)}{H} \right)}\\\cos{\left(\frac{\pi \left(- H/2 + v\right)}{H} \right)} \cos{\left(\frac{\pi \left(- W + 2 u\right)}{W} \right)}\end{matrix}\right]\right)
\end{equation} where $H$ and $W$ denote the resolution of the panorama. During $\alpha$-blending, we replace $\mathcal{G}^{\text{2D}}$ in Eqn.~\eqref{eq:alpha-blending} with the value of the tangent-plane Gaussian at $\mathbf{x}_{\text{2D}}$. Additionally, to render the depth of the panorama, we use the Euclidean distance from the primitive to the camera center in place of $\mu_z$. Finally, the $i$-th probe at position $\mathbf{p}_i$ stores a color panorama $\Phi$ and a depth panorama $\Theta$, enabling color $\mathbf{c}_i$ and depth $t_i$ queries via the direction $\mathbf{d}_i$ respectively:
\begin{equation}
    \mathbf{c}_i, t_i=\Phi\left({\mathbf{p}_i},\mathbf{d}_i\right),\Theta\left({\mathbf{p}_i},\mathbf{d}_i\right).
\label{eq:probes_query}
\end{equation} 

\begin{table}
  \caption{\textbf{List of important notations in GaussProbe and IterQuery.}}
  \label{tab:notation}
  \resizebox{\linewidth}{!}{
  \begin{tabular}{cl}
    \toprule
    \textbf{Symbol} & \textbf{Definition} \\
    \midrule
    $\boldsymbol{\mu}^{'},\mu_x,\mu_y,\mu_z$ & Position of 3D Gaussians in the camera space \\
    $\varphi$ & Optimal projection function\\
    $K, i$ & Number and index of Gaussian light field probes \\
    $\Phi, \Theta$ & Color and depth panorama of GaussProbe \\
    $\mathbf{p}_i, \mathbf{d}_i$ & Position and direction of $i$-th GaussProbe \\
    $\mathbf{c}_i, t_i$ & Color and depth of $i$-th GaussProbe queried by $\mathbf{d}_i$\\
    $\mathbf{o}+t\mathbf{d}$ & Queried ray (e.g., refracted or reflected ray) \\
    $\mathbf{o}+\hat{t}\mathbf{d}$ & Estimated intersection of the queried ray with Env. \\
    $w_i$ & Trilinear interpolation weight of $i$-th GaussProbe \\
\bottomrule
\end{tabular}
}
\end{table}
\begin{figure*}[!ht]
  \newlength{\lenSampleFF}
  \setlength{\lenSampleFF}{0.99\linewidth}
  \addtolength{\tabcolsep}{-4pt}
  \renewcommand{\arraystretch}{0.5}
  \centering
  \begin{tabular}{cccc}
  \includegraphics[width=0.25\lenSampleFF]{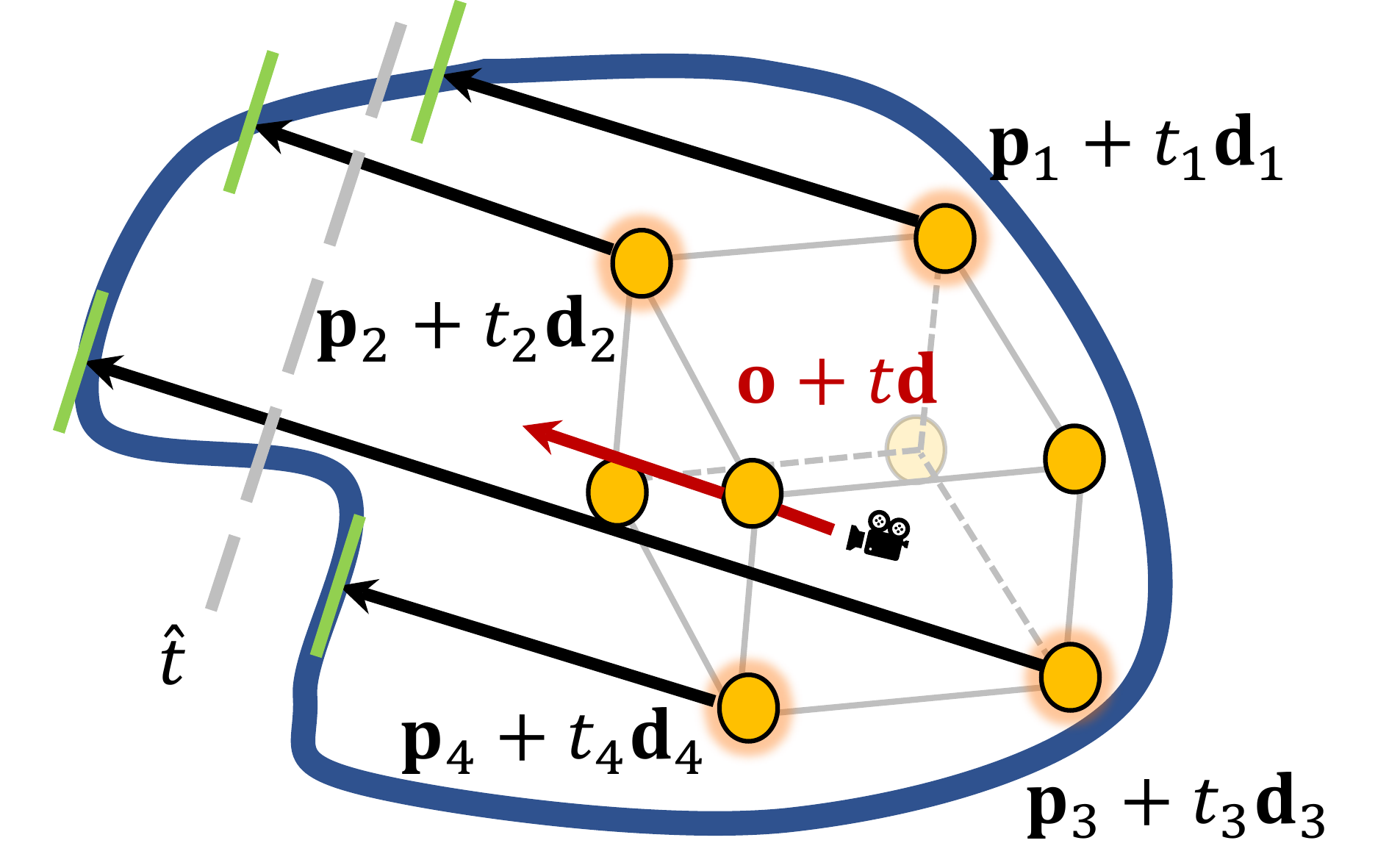} &
  \includegraphics[width=0.25\lenSampleFF]{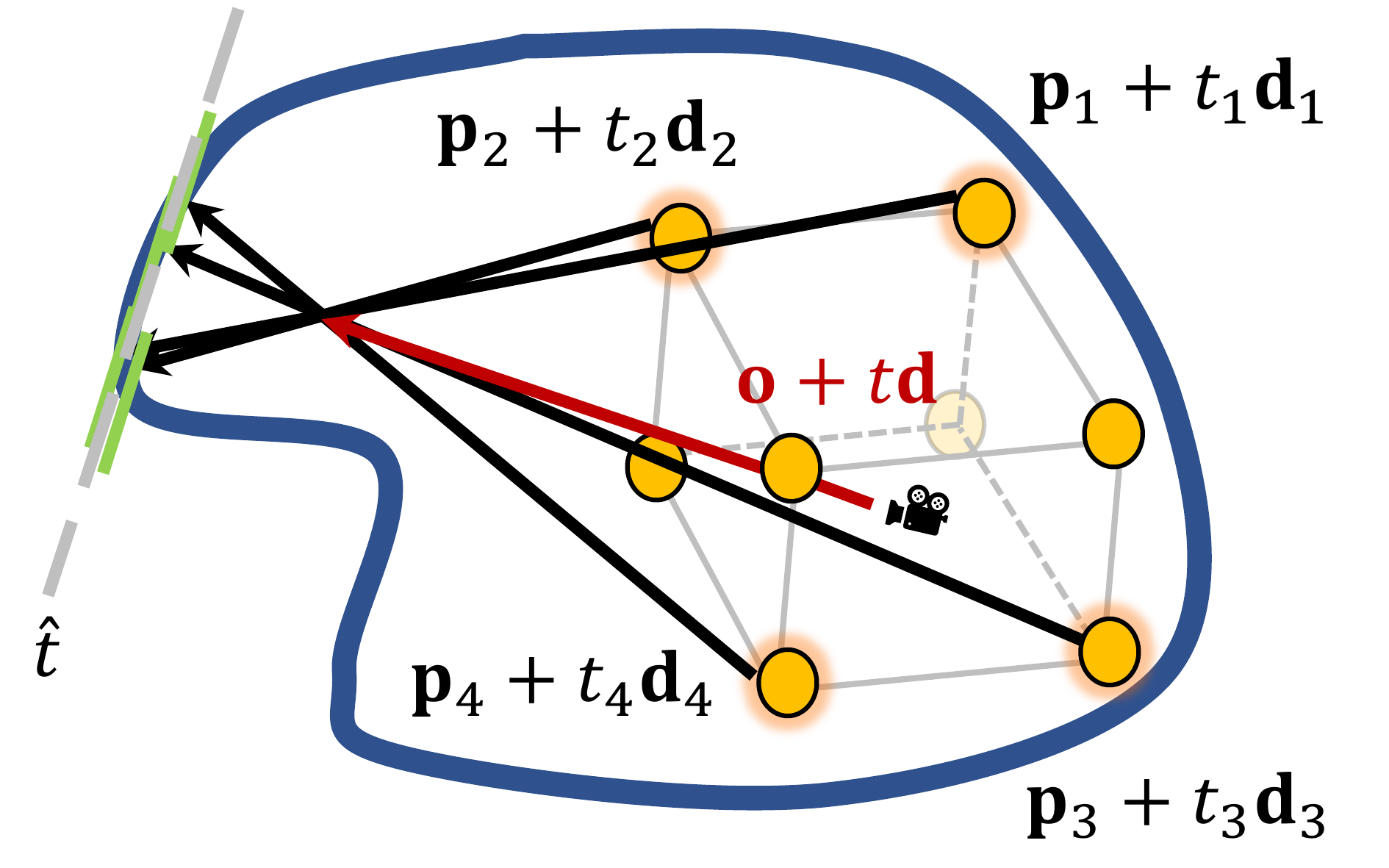} &
  \includegraphics[width=0.25\lenSampleFF]{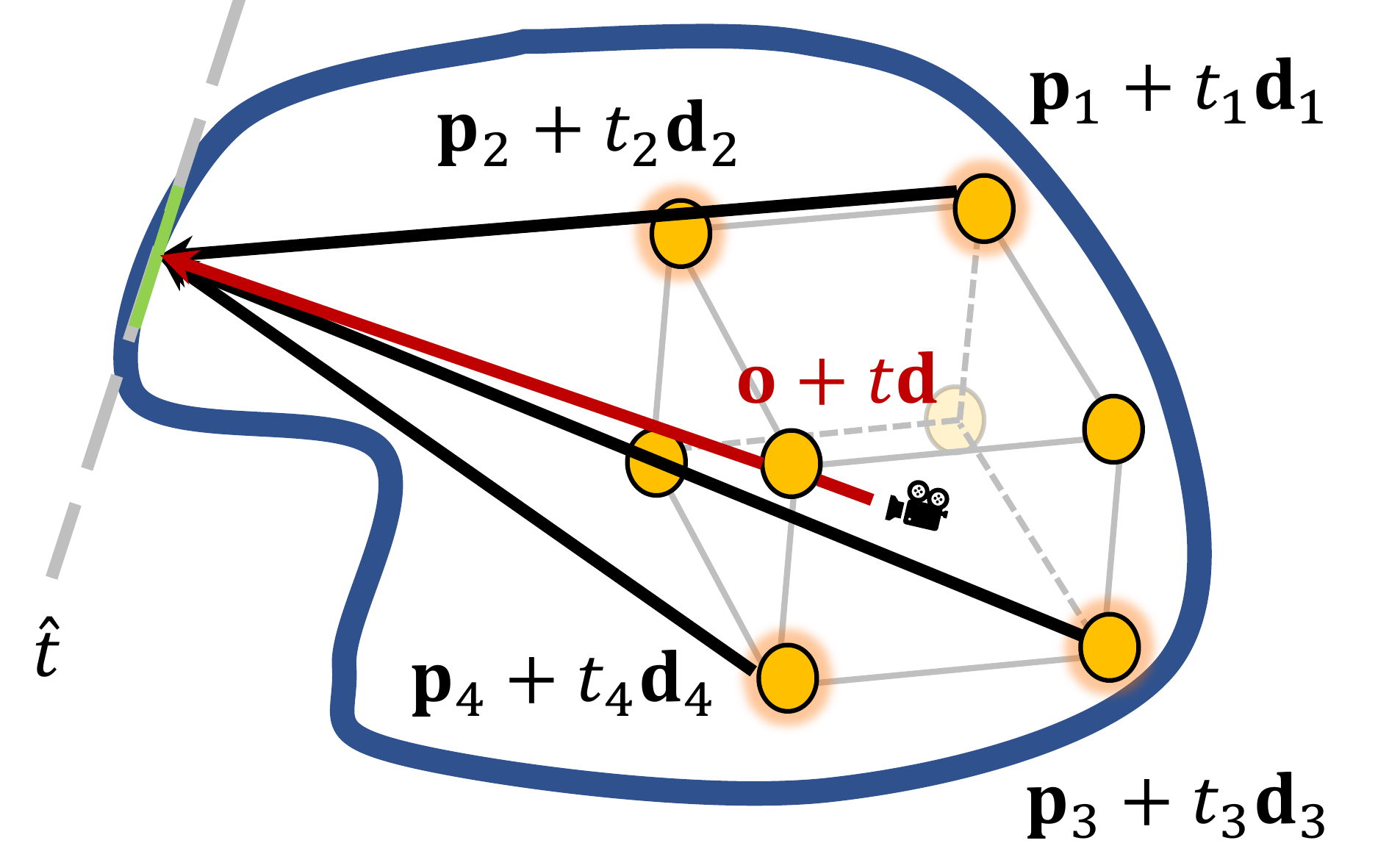} &
  \includegraphics[width=0.25\lenSampleFF]{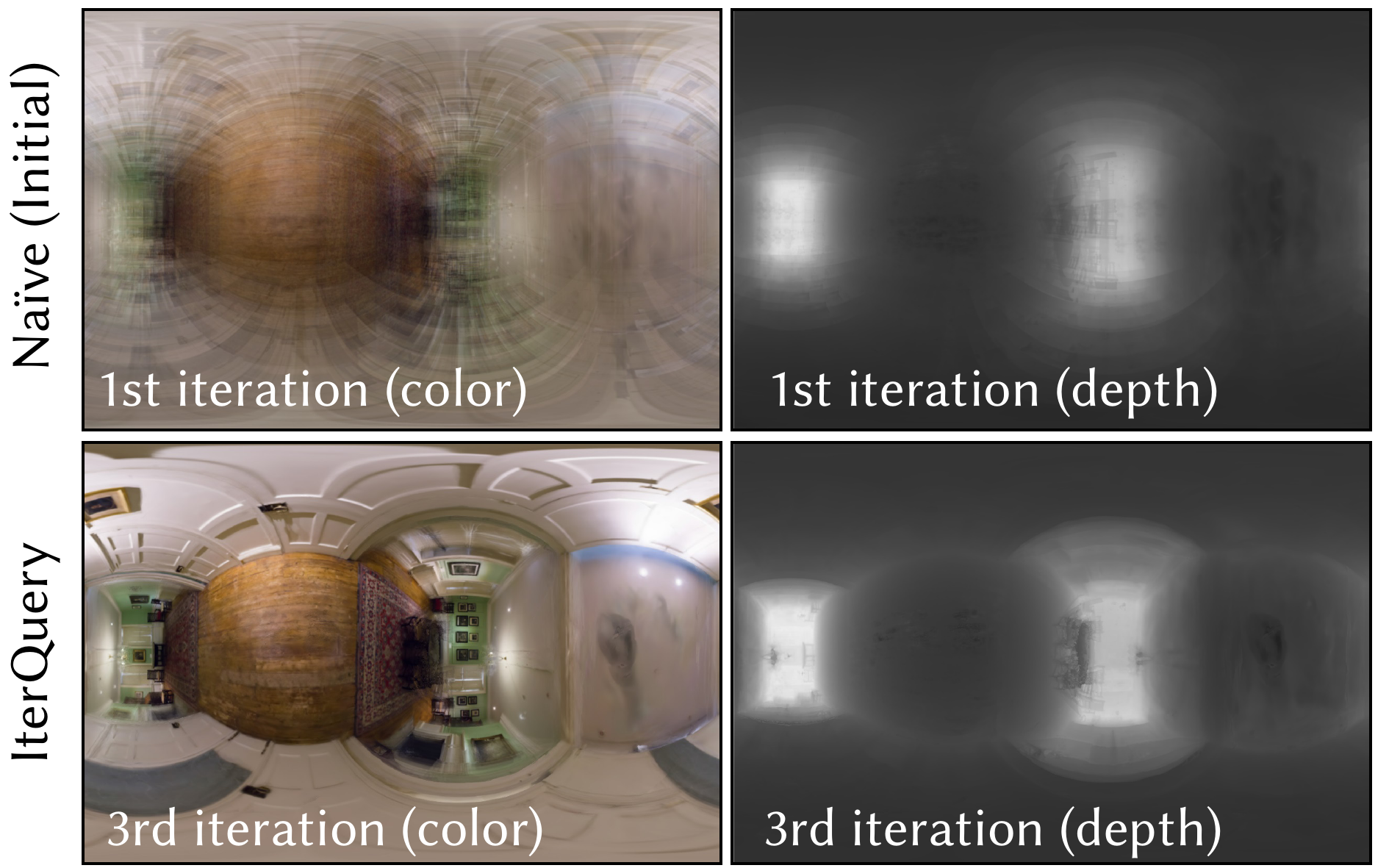} \\
  \raisebox{12pt}{{\small{ \textsf{(a) 1st iteration}}}} &
\raisebox{12pt}{{\small{ \textsf{(b) 2nd iteration}}}} &
  \raisebox{12pt}{{\small{ \textsf{(c) 3rd iteration }}}} &
    \raisebox{12pt}{{\small{ \textsf{(d) impact of IterQuery }}}}
  \end{tabular}
  \vspace{-0.2cm}
  \caption{\label{fig:probes_query}
\textbf{Illustration of our depth-based iterative Gaussian probes query (IterQuery) algorithm.}  For clarity, we illustrate four selected probes, with positions \(\mathbf{p}_1\), \(\mathbf{p}_2\), \(\mathbf{p}_3\), and \(\mathbf{p}_4\), chosen from the eight (orange circles). The black camera with the red arrow represents the queried ray (refracted or reflected ray), expressed as \(\mathbf{o} + t \mathbf{d}\). The goal of the algorithm is to determine the correct \( \hat{t} \), such that \(\mathbf{o} + \hat{t} \mathbf{d}\) corresponds to the first intersection between the queried ray and the scene, as well as the correct directions \(\mathbf{d_1}, \mathbf{d_2}, \mathbf{d_3}, \mathbf{d_4}\).
  }
\end{figure*}
\begin{figure}[tb]
\begin{pseudocode}[label={list:probes_query}, title={Pseudocode for our IterQuery algorithm.}]
def (*@\textbf{\textcolor{solarized_blue}{iter\_query}}@*)(probes, ray, total_iter):
    o, d = ray[..., :3], ray[..., 3:]
    # calculate the parallax relative to the virtual camera
    delta_depth = (*@\textbf{\textcolor{solarized_blue}{dot}}@*)(o - probes.pos, d) 
    d_i = d # initialize the directions of probes, (*@\textcolor{solarized_cyan}{\autoref{eq:initialize}}@*)
    while total_iter > 0:
        # (*@\textcolor{solarized_cyan}{ \autoref{eq:depth_query}}@*) and the green lines in (*@\textcolor{solarized_cyan}{\autoref{fig:probes_query} }@*)
        depth = probes.depth_map.(*@\textbf{\textcolor{solarized_blue}{texture}}@*)(d_i) 
        cos_depth = depth * (*@\textbf{\textcolor{solarized_blue}{dot}}@*)(d, d_i) + delta_depth
        # (*@\textcolor{solarized_cyan}{\autoref{eq:tri_lerp}}@*) and the gray dashed lines  in (*@\textcolor{solarized_cyan}{\autoref{fig:probes_query}}@*)
        t_hat = (*@\textbf{\textcolor{solarized_blue}{tri\_lerp}}@*)(cos_depth, o, probes.pos)
        d_i = (*@\textbf{\textcolor{solarized_blue}{normalize}}@*)(o + t_hat * d - probes.pos) # (*@\textcolor{solarized_cyan}{\autoref{eq:probes_update}}@*)
        if abs(depth - probes.depth_map.(*@\textbf{\textcolor{solarized_blue}{texture}}@*)(d_i)) <= (*@$\epsilon$@*):  
            break # exit if converged in a certain region
        total_iter -= 1
    color = probes.color_map.(*@\textbf{\textcolor{solarized_blue}{texture}}@*)(d_i)
    color = (*@\textbf{\textcolor{solarized_blue}{tri\_lerp}}@*)(color, o, probes.pos)
    return color
\end{pseudocode}
\end{figure}

\paragraph{Depth-based iterative probes query} When computing refracted or reflected light during inverse rendering, we should query probes with respect to a given ray. Let \(\mathbf{o}\) and \(\mathbf{d}\) denote the origin and direction of the queried ray, respectively. We also denote \(\mathbf{p}_i\) and \(\mathbf{d}_i\) as the position and direction of the \(i\)-th probe. Fig.~\ref{fig:probes_query}(d) illustrates that naively querying $K$ probes along the same direction $\mathbf{d}_i=\mathbf{d}$ and averaging the results leads to over-blurriness due to the parallax among the probes. To better capture details in specular reflection and refraction, we propose a depth-based iterative probes query (IterQuery) algorithm tailored for our baked GaussProbe. As depicted in Fig.~\ref{fig:probes_query}, we initialize the directions of probes $\mathbf{d}_i$ with the direction of the queried ray $\mathbf{d}$:
\begin{equation}
    \mathbf{d}_i:=\mathbf{d}.
    \label{eq:initialize}
\end{equation} Then we query the depth $t_i$ to obtain the intersection points $\mathbf{p}_i+t_i\mathbf{d}_i$ of the $i$-th probe with the environment:
\begin{equation}
    t_i=\Theta\left({\mathbf{p}_i}, \mathbf{d}_i\right).
\label{eq:depth_query}
\end{equation} 
The trilinear interpolation of the depth maps projected onto the queried ray is given by:
\begin{equation}
\label{eq:tri_lerp}
\begin{split}
\hat{t}&=\sum_{i=1}^{K}w_i\left(\left(\mathbf{p}_i+t_i\mathbf{d}_i-\mathbf{o}\right)\cdot\mathbf{d}\right)
\end{split}
\end{equation} where $\cdot$ denotes the dot product of vectors and \(w_i\) is the interpolation weights. Then $\mathbf{d}_i$ are updated using the following formulation:
\begin{equation}
\mathbf{d}_i:=\frac{\mathbf{o}+\hat{t}\mathbf{d}-\mathbf{p}_i}{\left\Vert{\mathbf{o}+\hat{t}\mathbf{d}-\mathbf{p}_i}\right\Vert}.
\label{eq:probes_update}
\end{equation} Then, the updated $\mathbf{d}_i$ are used to query in the same manner as the previous iteration. The depth queried from the probe (green dashed lines) corresponds to a point on the surface of the 3D scene, while 
  $\hat{t}$ corresponds to a point on the ray to be queried (gray dashed lines). Therefore, when \( \hat{t} \) remains constant, the ray to be queried should intersect with the scene. We provide a simplified formal proof for the case where $K = 1$. Since the convergence condition requires \( \hat{t} \) to remain constant, we can simultaneously solve Eqn.~\eqref{eq:depth_query}, Eqn.~\eqref{eq:tri_lerp} and Eqn.~\eqref{eq:probes_update} to derive the following formulation:
\begin{gather}
\hat{t}=\left(\mathbf{p}_i+\frac{\mathbf{o}+\hat{t}\mathbf{d}-\mathbf{p}_i}{\left\Vert{\mathbf{o}+\hat{t}\mathbf{d}-\mathbf{p}_i}\right\Vert}\Theta\left(\mathbf{p}_i, \mathbf{d}_i\right)-\mathbf{o}\right)\cdot\mathbf{d}\\
\implies \Theta\left(\mathbf{p}_i, \mathbf{d}_i\right)=t_i=\left\Vert{\mathbf{o}+\hat{t}\mathbf{d}-\mathbf{p}_i}\right\Vert\\
\implies\mathbf{o}+\hat{t}\mathbf{d}=\mathbf{p}_i+t_i\mathbf{d}_i. 
\end{gather} After the iteration stops, the final color can be obtained by replacing \(\Theta\) with \(\Phi\) in Eqn.~\eqref{eq:tri_lerp}. The pseudocode describing this algorithm is provided in \autoref{list:probes_query}.

Theoretically, the time complexity of the IterQuery algorithm is $\mathcal{O}\left(TKQ\right)$. $T$, $K$, and $Q$ denote the total number of iterations, the number of probes, and the number of queried rays, respectively. The required number of iterations is influenced by the number of probes $K$ and the depth panorama $\Theta$, which we discuss further in the experimental section. Since only a few iterations are needed to achieve a significant improvement compared to the 1st iteration, our approach is both efficient and high-quality. 

\begin{figure}[!t]
  \centering
  \includegraphics[width=0.98\linewidth]{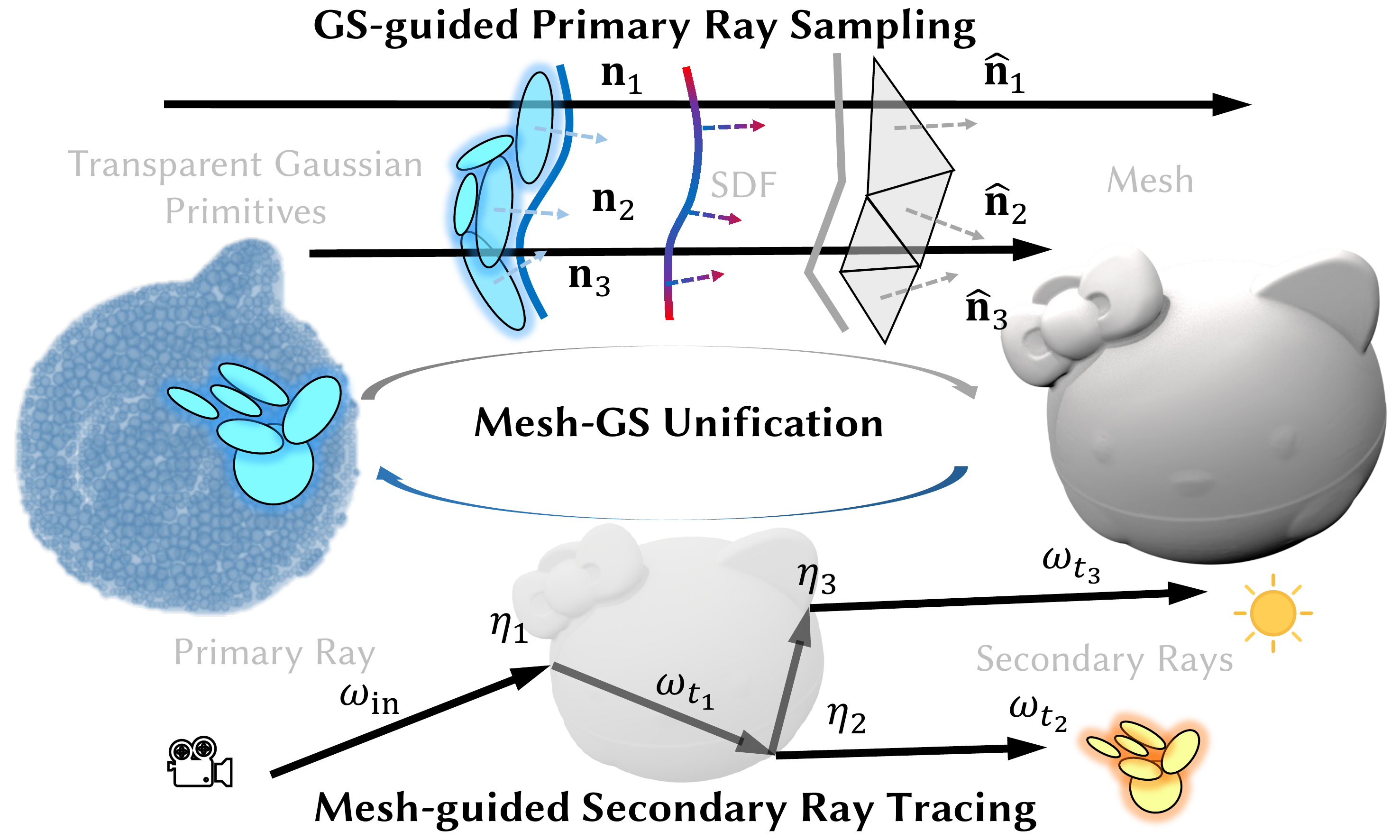} 
  \caption{\label{fig:gsmesh}
    \textbf{Illustration of our mesh-GS fusion strategy. }   \textbf{GS}$\rightarrow$\textbf{Mesh: } We use transparent Gaussian primitives to guide the primary ray sampling of the SDF, efficiently generating an accurate mesh. \textbf{Mesh}$\rightarrow$\textbf{GS: } We employ the mesh as a proxy for fast secondary ray tracing, which are then utilized in the IterQuery.
  }
\end{figure}

\subsection{Multi-Stage Reconstruction}

Similar to previous works~\cite{liu2023nero, liang2024gs}, we adopt the multi-stage reconstruction strategy to achieve the inverse rendering. The first stage involves environment reconstruction and baking, where the baked GaussProbe are used for light computations during geometry and material reconstruction. In the second stage, we leverage transparent Gaussian primitives to reconstruct geometry and material in a physically-based deferred rendering pipeline. During the process, we unify the explicit mesh and GS to trace secondary rays, effectively refining inverse rendering.

\paragraph{Optimization and losses} During the process of the multi-stage reconstruction, we impose some regularization terms on the final loss. For the normal regularization term, we follow the previous works~\cite{wang2023nemto, jiang2024gaussianshader}:
\begin{equation}
    \mathcal{L}_{\text{normal}}=1-\mathcal{N}\cdot\hat{\mathcal{N}_{\mathcal{D}}}
\end{equation} where $\mathcal{N}$ denotes the rendered normal map, and $\hat{\mathcal{N}_{\mathcal{D}}}$ is the gradient of the rendered depth map. We also incorporate the mask term and the D-SSIM term into the total loss $\mathcal{L}$, i.e.,
\begin{equation}
    \mathcal{L}=(1-\lambda_1)\mathcal{L}_1+\lambda_1\mathcal{L}_{\text{D-SSIM}}+\lambda_2\mathcal{L}_{\text{normal}}+\lambda_3\mathcal{L}_{\text{mask}}
\end{equation} where $\mathcal{L}_{1}$ and $\mathcal{L}_{\text{D-SSIM}}$ are the rendering losses.

\paragraph{Unifying mesh and GS for secondary ray effects} Physically-based refraction produces at least two bounces of rays inside transparent objects. The rasterization-based 3D-GS~\cite{kerbl20233d}, however, is unable to handle secondary rays. Moreover, the second-bounce ray marching is known to be expensive for previous NeRF-based methods~\cite{gao2023transparent, li2024tensosdf}. While using an MLP to predict the refracted direction~\cite{wang2023nemto} or light~\cite{sun2024nu} is effective, it tends to lose details. Inspired by the joint optimization of GS and SDF~\cite{yu2024gsdf, wu2024deferredgs}, as well as the unification of SDF and mesh~\cite{li2024tensosdf}, we unify the transparent Gaussian primitives and mesh to address this issue. As detailed in Fig.~\ref{fig:gsmesh}, during the reconstruction, we leverage the hitting points map  $\mathcal{X}$ rendered by transparent Gaussian primitives as guidance for the primary ray sampling of the SDF. The rendered normal map $\mathcal{N}$ is used to regularize the gradients, which improves the accuracy of the estimated SDF. We extract the explicit mesh from
the SDF field via a marching cube algorithm~\cite{lorensen1987marching}. On the other hand, we employ the extracted mesh as a proxy to efficiently trace secondary rays for subsequent GaussProbe queries. Our mesh-GS unification is somewhat similar to SuGaR~\cite{guedon2024sugar}, which jointly optimizes the mesh and 3D Gaussians located on the surface of the mesh. However, our method focuses more on leveraging the fast reconstruction capability of 3D-GS~\cite{kerbl20233d} and the efficiency of mesh in tracing secondary rays to further refine inverse rendering. 

\begin{figure}[!t]
  \centering
  \includegraphics[width=0.92\linewidth]{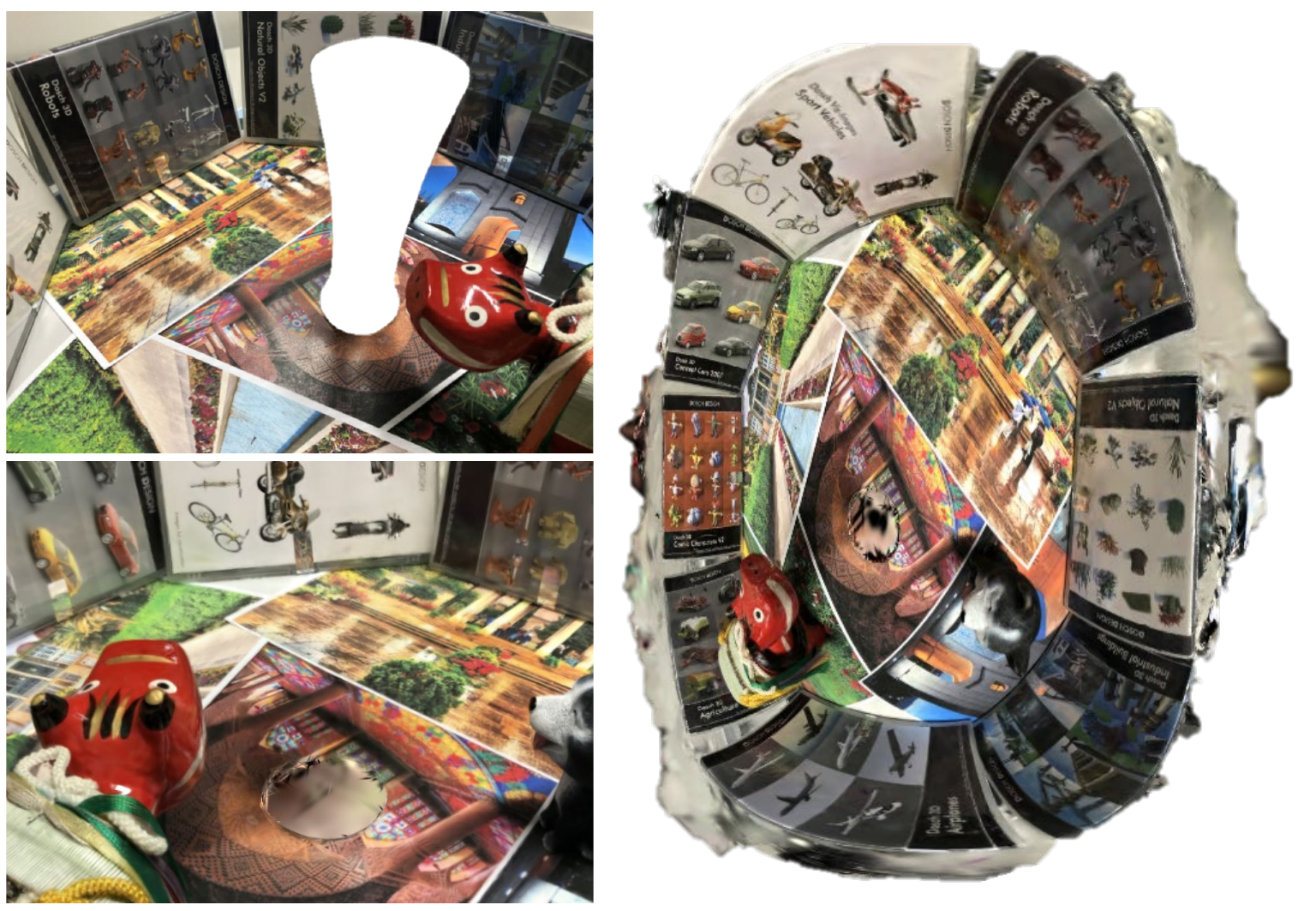} 
  \caption{\label{fig:sam}
    \textbf{3D scene segmentation results on the \textsc{Glass} scene. }Top left: Image segmentation results. Bottom left: Segmented scene represented by the original 3D-GS~\cite{kerbl20233d}. Right: Probes baked from the segmented scene.
  }
\end{figure}
\begin{figure*}[!t]
  \centering
  \includegraphics[width=1\linewidth]{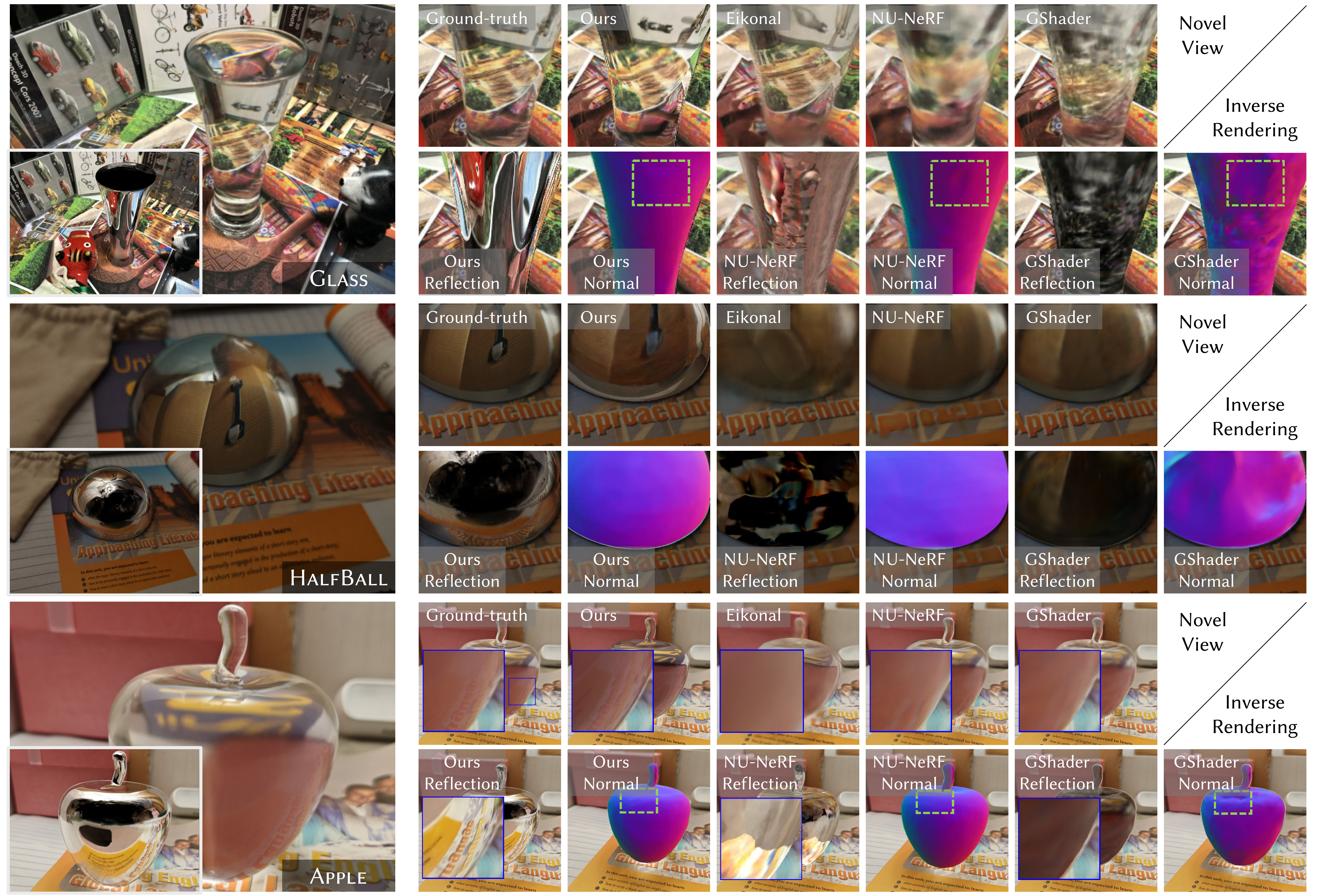} 
  \caption{\label{fig:comp}
    \textbf{\textbf{Qualitative comparison of novel-view synthesis and inverse rendering results on real-captured scenes with colorless transparent objects.}}  The left image showcases the transparent object along with its specular reflection component decoupled using our method. The black regions in the reflection component represent the areas that are not visible in the dataset.}
\end{figure*}

\begin{figure*}[!t]
  \centering
  \includegraphics[width=1\linewidth]{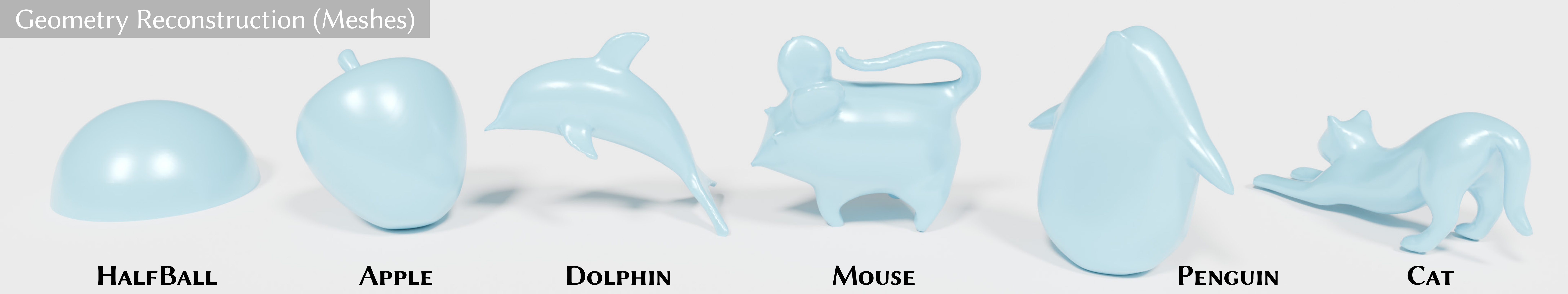} 
  \caption{\label{fig:meshes_results}
    \textbf{Surface mesh reconstruction results of our method on the real-captured and synthetic datasets.}  } 
\end{figure*}
\begin{figure*}[!t]
  \centering
  \includegraphics[width=1\linewidth]{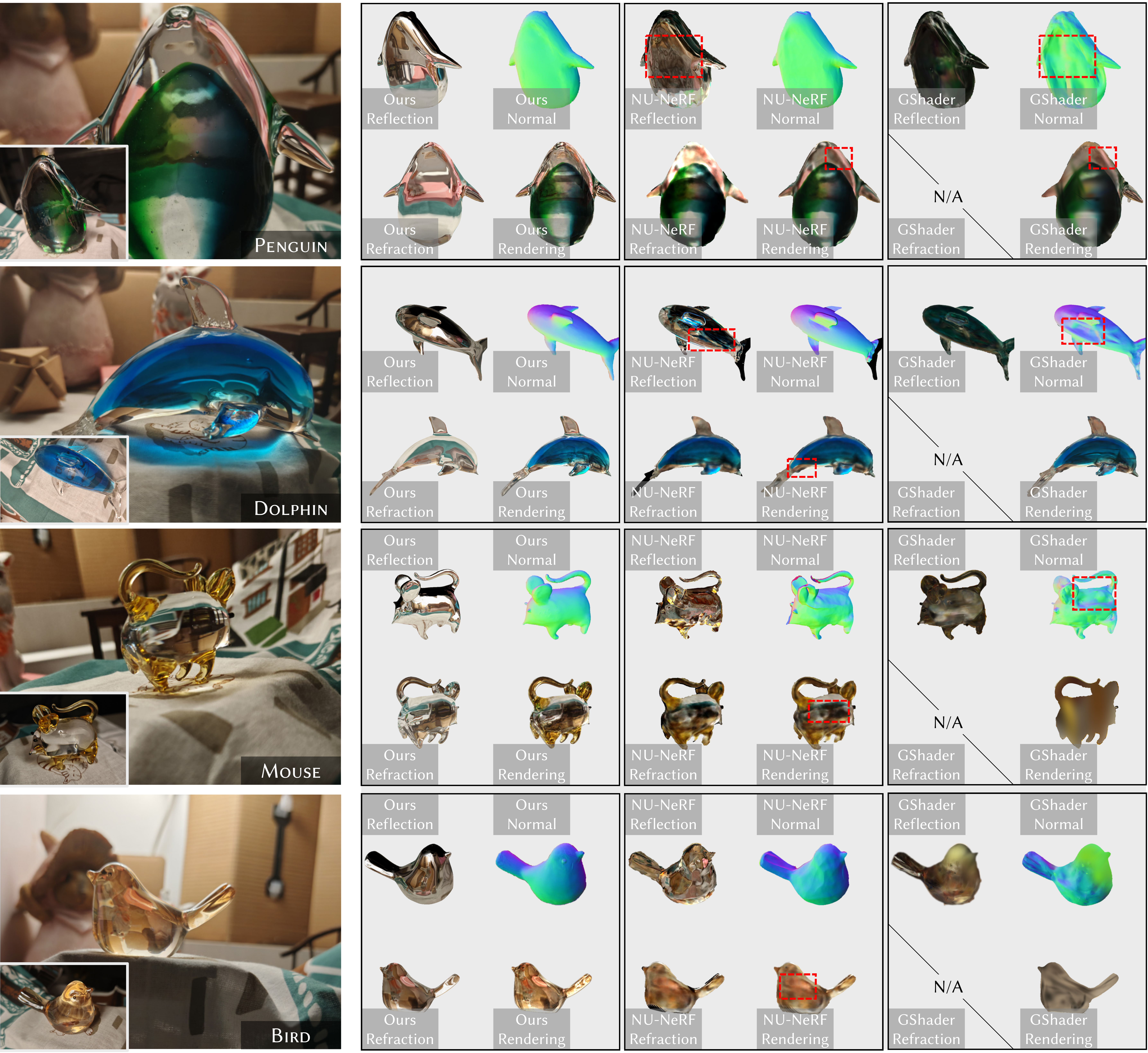} 
  \caption{\label{fig:comp_color}
    \textbf{\textbf{Qualitative comparison of novel-view synthesis and inverse rendering results on real-captured scenes with colored transparent objects.}} The left images showcase the input images of the real-captured scenes with colored transparent objects. By handling absorption as described in Eqn.~\eqref{eq:handling_color}, we can effectively decouple the refracted light from the objects' inherent texture, whereas NU-NeRF~\cite{sun2024nu} can only yield coupled results.}
    \vspace{-0.4cm}
\end{figure*}
\begin{figure*}[!t]
  \centering
  \includegraphics[width=1\linewidth]{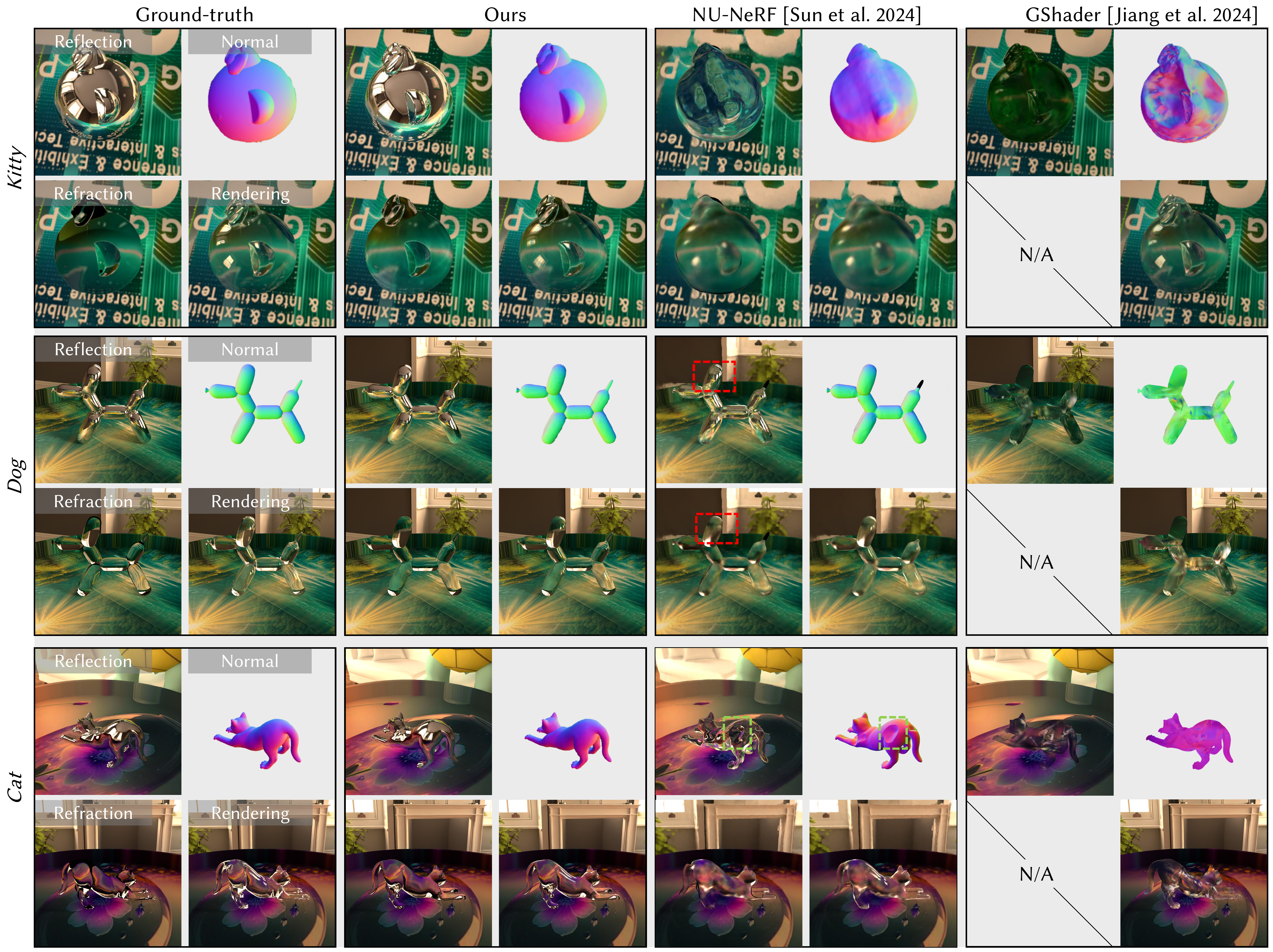} 
  \caption{\label{fig:comp_synthetic}\textbf{Qualitative comparison of novel-view synthesis and inverse rendering results on the synthetic dataset.}}
\end{figure*}

\section{Experiments}

\subsection{Experiment Settings}

\paragraph{Implementation details} We implement our method based on the PyTorch framework in 3D-GS~\cite{kerbl20233d} with the Adam optimizer~\cite{kingma2014adam}. Currently, for the losses, we set $\lambda_1=0.2, \lambda_2=0.2, \lambda_3=1$ in all the tests. We employ the pre-trained SAM2~\cite{ravi2024sam, kirillov2023segment} guided by GroundingDINO~\cite{liu2025grounding} for the 3D scene segmentation, as shown in Fig.~\ref{fig:sam}. Specifically, by entering text prompts, in conjunction with RGB images, GroundingDINO can generate bounding boxes for the transparent object. Subsequently, these bounding boxes can be utilized as box prompts for SAM2 to produce segmentation results. These results act as regularization for Gaussians, facilitating 3D scene segmentation. For our Gaussian light field probes, we set the iteration count of the IterQuery algorithm to 5 and empirically set the number of Gaussian probes to 8 or 64. For all baseline methods, we adopt their official implementations. 

\begin{figure*}[!t]
  \centering
  \includegraphics[width=1\linewidth]{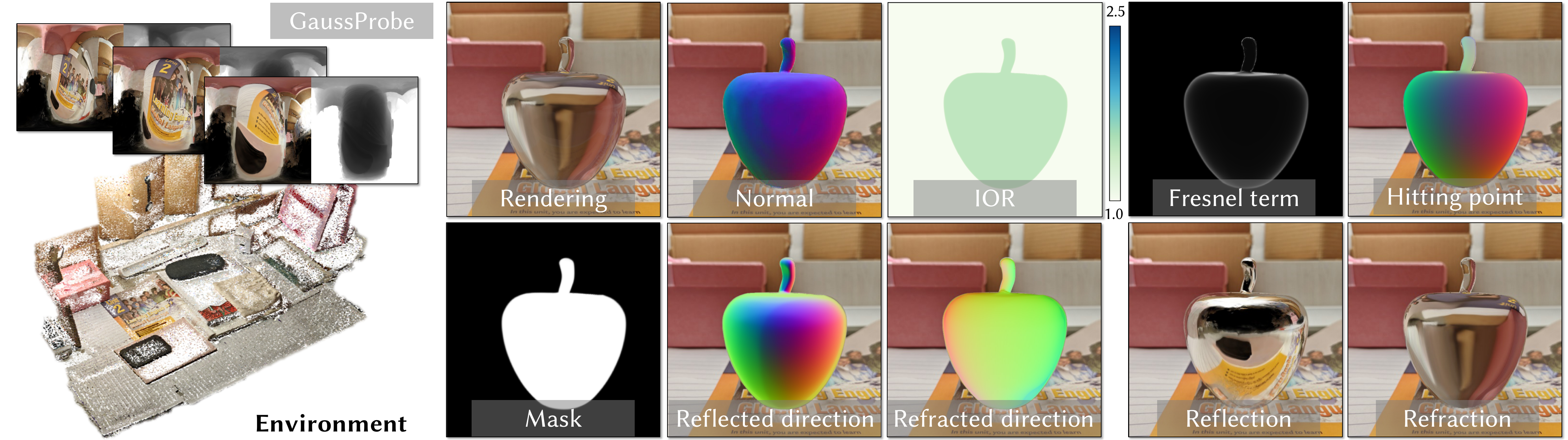} 
  \caption{\label{fig:intermediate} 
  \textbf{Detailed intermediate results of our method.} Left: the environment and the corresponding GaussProbe. Right: maps of additional parameters.}
\end{figure*}
\newcommand{\winner}[1]{\fontseries{b}\selectfont{#1}}
\newcommand{\scene}[1]{\textsc{#1}}
\begin{table}[!t]
    \setlength{\tabcolsep}{0.2pt} 
    \centering
    \caption{
    \textbf{Quantitative comparison of novel view synthesis results on the real-captured datasets with colorless transparent objects.}
    }
    \vspace{-2ex}
    \label{tab:comparison}
\resizebox{\linewidth}{!}{
    \begin{tabular}{lccccccccc
        }
        \toprule
        &
        \multicolumn3c{\scene{Glass}}&
        \multicolumn3c{\scene{HalfBall}}&
        \multicolumn3c{\scene{Apple}}
        \\
        \cmidrule(lr){2-4}
        \cmidrule(lr){5-7}
        \cmidrule(lr){8-10}
        \multirow{-2}{*}{Methods} 
        &
        \multicolumn1c{PSNR$\uparrow$}&
        \multicolumn1c{SSIM$\uparrow$}&
        \multicolumn1c{LPIPS$\downarrow$}&
        \multicolumn1c{PSNR$\uparrow$}&
        \multicolumn1c{SSIM$\uparrow$}&
        \multicolumn1c{LPIPS$\downarrow$}&
        \multicolumn1c{PSNR$\uparrow$}&
        \multicolumn1c{SSIM$\uparrow$}&
        \multicolumn1c{LPIPS$\downarrow$}
        \\
        \midrule        
Eikonal & \bfseries 27.15 & 0.941 &  0.057  &  27.43 & 0.890 & 0.144  &  20.47 & 0.951 & 0.067  \\
GShader & 26.52  & 0.951 & 0.052  & 27.29 & 0.953 & 0.096  & 20.97 & 0.955 & 0.068  \\
NU-NeRF & 26.78 & 0.942& 0.071  & 27.48 & 0.946& 0.149 & 22.25 &0.963 & 0.057\\

Ours & 27.12 & \bfseries 0.952 &  \bfseries 0.044  &  \bfseries 28.07 & \bfseries 0.954 & \bfseries 0.084  & \bfseries 23.05 & \bfseries 0.965 & \bfseries 0.047  \\
        \bottomrule
    \end{tabular}
    }
\end{table}

\begin{table}[!t]
    \setlength{\tabcolsep}{0.2pt} 
    \centering
    \caption{
    \textbf{Quantitative comparison of novel view synthesis results on the real-captured datasets with colored transparent objects.}
    }
    \vspace{-2ex}
    \label{tab:comparison_color}
\resizebox{\linewidth}{!}{
    \begin{tabular}{lccccccccc
        }
        \toprule
        &
        \multicolumn3c{Ours}&
        \multicolumn3c{GShader}&
        \multicolumn3c{NU-NeRF}
        \\
        \cmidrule(lr){2-4}
        \cmidrule(lr){5-7}
        \cmidrule(lr){8-10}
        \multirow{-2}{*}{Scenes} 
        &
        \multicolumn1c{PSNR$\uparrow$}&
        \multicolumn1c{SSIM$\uparrow$}&
        \multicolumn1c{LPIPS$\downarrow$}&
        \multicolumn1c{PSNR$\uparrow$}&
        \multicolumn1c{SSIM$\uparrow$}&
        \multicolumn1c{LPIPS$\downarrow$}&
        \multicolumn1c{PSNR$\uparrow$}&
        \multicolumn1c{SSIM$\uparrow$}&
        \multicolumn1c{LPIPS$\downarrow$}
        \\
        \midrule        
\scene{Penguin} & 22.09 & \bfseries 0.832 &  \bfseries 0.255  &  21.51 & 0.806 & 0.271  & \bfseries 22.29 & 0.821 & 0.318 \\

\scene{Dolphin} & 22.66 & 0.832 &  \bfseries 0.142  &  22.60 & 0.842 & 0.158  & \bfseries 22.70 & \bfseries 0.833 & 0.306  \\
\scene{Mouse} & \bfseries 20.37 & \bfseries 0.695 &  \bfseries 0.154  &  19.53 & 0.686 & 0.211 & 19.86 & 0.673 & 0.300  \\
\scene{Bird} & 21.09 & 0.830 &  \bfseries 0.136  & \bfseries 21.30 & \bfseries 0.832 & 0.161  & 21.00 & 0.820 &  0.306 \\
        \midrule
        Average & \bfseries 21.55 & \bfseries 0.797 & \bfseries 0.172 & 21.24 & 0.792 & 0.200 & 21.46 & 0.787 & 0.308 \\
        \bottomrule
    \end{tabular}
    }
\end{table}

\begin{table}[!t]
    \setlength{\tabcolsep}{2pt} 
    \centering
    \caption{
    \textbf{Quantitative comparison of novel view synthesis and inverse rendering results on the synthetic dataset.} The refraction term of GShader~\cite{jiang2024gaussianshader} is marked as “N/A” due to its its incapability of modeling refracted light.
    }
    \vspace{-2ex}
    \label{tab:comparison_synthetic}
\resizebox{\linewidth}{!}{
    \begin{tabular}{lccccccc
        }
        \toprule
        &
        \multicolumn3c{Novel View Synthesis}&
        \multicolumn1c{Normal}&
        \multicolumn1c{Reflection}&
        \multicolumn1c{Refraction}&
        \multicolumn1c{Base Color}
        \\
        \cmidrule(lr){2-4}
        \cmidrule(lr){5-5}
        \cmidrule(lr){6-6}
        \cmidrule(lr){7-7}
        \cmidrule(lr){8-8}
        \multirow{-2}{*}{Methods} 
        &
        \multicolumn1c{PSNR$\uparrow$}&
        \multicolumn1c{SSIM$\uparrow$}&
        \multicolumn1c{LPIPS$\downarrow$}&
        \multicolumn1c{MAE$^{\circ}\downarrow$}&
        \multicolumn1c{PSNR$\uparrow$}&
        \multicolumn1c{PSNR$\uparrow$}&
        \multicolumn1c{PSNR$\uparrow$}
        \\
        \midrule        
GShader & 24.05 & 0.922 & 0.069  &  26.51 & 12.44 & N/A  & 13.19 \\
NU-NeRF & 22.52 & 0.759 & 0.266 & 16.02 & 13.65 & 19.90 & 17.08 \\

Ours & \bfseries 25.66 & \bfseries 0.935 &  \bfseries 0.064  &  \bfseries 5.53 & \bfseries 17.60 & \bfseries 22.87  & \bfseries 21.51 \\
        \bottomrule
    \end{tabular}
    }
\end{table}

\paragraph{Datasets} We conduct evaluation on several datasets with transparent
objects, including synthetic datasets and real-captured datasets. Specifically, we take one scene including a refractive object with unknown geometry from~\citet{bemana2022eikonal}: \textsc{Glass}, and capture 6 scenes by ourselves: \textsc{HalfBall}, \textsc{Apple}, \textsc{Dolphin}, \textsc{Penguin}, \textsc{Mouse} and \textsc{Bird}. We use a smartphone camera to capture $80\sim200$ views for each scene. To demonstrate that our method can handle indirect light from nearby contents beyond opaque planes~\cite{gao2023transparent}, we place several geometrically and texturally complex objects around the transparent objects. Additionally, the last 4 scenes are designed to validate our modeling of colored transparent objects. To qualitatively compare the decoupling and reconstruction quality of our method with other approaches, we render a synthetic dataset with ground truth maps using the Blender: \textsc{Kitty}, \textsc{Dog} and \textsc{Cat}.

\paragraph{Baselines and metrics} We compare our method against the following baselines: 
\begin{itemize}
    \item \textbf{GShader}~\cite{jiang2024gaussianshader}: an inverse rendering framework for glossy objects based on 3D-GS~\cite{kerbl20233d}.
    \item \textbf{Eikonal}~\cite{bemana2022eikonal}: the  state-of-the-art (SOTA) novel view synthesis method for refractive objects based on NeRF~\cite{mildenhall2020nerf}.
    \item \textbf{NU-NeRF}~\cite{sun2024nu}: the SOTA reconstruction method for transparent objects based on SDF~\cite{wang2021neus, ge2023ref}.
\end{itemize} We do not compare with NEMTO~\cite{wang2023nemto} and \citet{gao2023transparent}, as they do not work with complex nearby objects and NU-NeRF has already been compared with both. For the novel view and the decoupled reflection, refraction, and base color components, we present quantitative results measured with three standard metrics: Peak Signal-to-Noise Ratio (PSNR), Structural Similarity Index (SSIM)~\cite{wang2004image} and Learned Perceptual Image Patch Similarity (LPIPS)~\cite{zhang2018unreasonable}. For the reconstructed normal maps, we use Mean Angular Error in degrees (MAE$^\circ$) to
evaluate the normal reconstruction accuracy.

\begin{figure}[!t]
  \centering
  \includegraphics[width=0.92\linewidth]{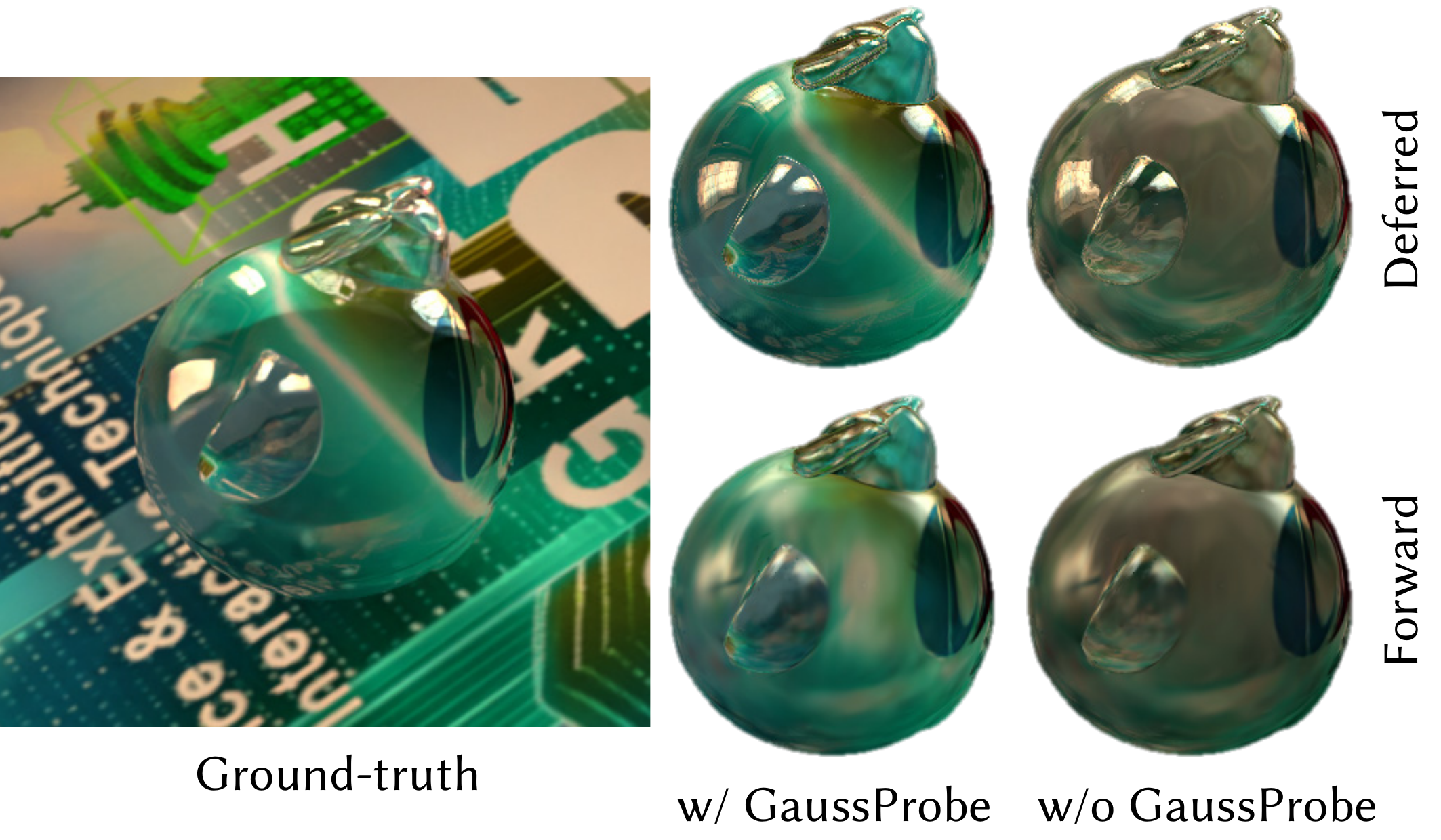} 
  \caption{\label{fig:exp_defer}
    \textbf{Qualitative ablation studies on the \textsc{KITTY} scene.} “Deferred” and “Forward" mean the deferred shading strategy and the forward shading strategy, respectively.}
\end{figure}
\begin{table}[!t]
    \setlength{\tabcolsep}{2.6pt} 
    \centering
    \caption{
    \textbf{Quantitative ablation studies on the \textsc{Kitty} scene.} “Deferred” and “Forward" mean the deferred shading strategy and the forward shading strategy, respectively.
    }
    \vspace{-2ex}
    \label{tab:ablation}
\resizebox{0.7\linewidth}{!}{
    \begin{tabular}{cc||ccc}
        \hline
        {Deferred}&
        {GaussProbe} &
        {PSNR$\uparrow$}&
        {SSIM$\uparrow$}&
        {LPIPS$\downarrow$}        \\
        \hline        
                \xmark & \xmark & 27.05 & 0.958 & 0.044\\
                \xmark & \cmark & 27.89 & 0.963 & 0.040\\
                \cmark & \xmark & 27.13 & 0.958 & 0.043 \\
                \cmark & \cmark & \bfseries 28.41 & \bfseries 0.970 & \bfseries 0.036\\
        \hline
    \end{tabular}
    }
\end{table}

\subsection{Results and Evaluation}

\paragraph{Qualitative comparison} In Fig.~\ref{fig:comp} and Fig.~\ref{fig:comp_color}, we evaluate our method on the real-captured dataset. And Fig.~\ref{fig:comp_synthetic} showcases the evaluation on the synthetic dataset. As seen, our method is capable of reconstructing transparent objects both effectively and efficiently. It is observed that in some cases even NU-NeRF~\cite{sun2024nu} has remaining artifacts that our method avoids, such as \textsc{Glass} and \textsc{Kitty}. And other methods~\cite{bemana2022eikonal, jiang2024gaussianshader} cannot faithfully generate normal maps due to the lack of surface representation or refraction modeling. Moreover, our method is adept at capturing the high-frequency details of specular reflection and refraction (e.g., the nearby contents in the inter-reflection - in \textsc{Apple} - or in the refraction - in \textsc{Mouse}). NU-NeRF~\cite{sun2024nu} tends to generate blurred results and fails to capture the detailed indirect light due to its light representation. It also should be noted that we can handle colored transparent objects and decouple the refraction term from the inherent color. Previous methods either fail to handle absorption or produce entangled results. We provide the surface mesh reconstruction results of these transparent objects, as shown in Fig.~\ref{fig:meshes_results}. The detailed intermediate results of our method are illustrated in Fig.~\ref{fig:intermediate}.

\paragraph{Quantitative comparisons} 
Tab.~\ref{tab:comparison} and Tab.~\ref{tab:comparison_color} show the quantitative comparison of novel view synthesis results on the real-captured datasets. We surpass other methods in most scenes, particularly in terms of LPIPS, and achieve the highest average metrics. Since real-captured scenes lack ground-truth geometry, material, and light, we also
evaluate our method on the synthetic dataset, as reported in Tab.~\ref{tab:comparison_synthetic}. We achieve state-of-the-art results in both novel view synthesis and inverse rendering. In terms of novel view synthesis on the synthetic dataset, the relatively high score of GShader~\cite{jiang2024gaussianshader} can be attributed to its spherical harmonic function that overfits the views, which also prevents it from reconstructing the correct normal.

\paragraph{Performance analysis} Generally, our method achieves superior inverse rendering quality compared to Eikonal~\cite{bemana2022eikonal} (22-24 hours) and NU-NeRF~\cite{sun2024nu} (8-9 hours), while requiring a comparable training time to GShader~\cite{jiang2024gaussianshader} (1 hour). We segment the 3D scene and bake GaussProbe within a few minutes. In terms of novel view synthesis and re-rendering performance, the runtime depends on the number of probes, the number of iterations, and the image resolution. For an \(800 \times 800\) image, the time cost is 0.002 seconds for a single probe with 1 iteration and 0.005 seconds for 8 probes with 5 iterations. Our transparent Gaussian primitives, similar to other 3D-GS-based methods~\cite{ye20243defer, jiang2024gaussianshader}, are capable of rendering various attributes maps in real-time for deferred shading. Due to the efficient designs, our method achieves the frame rate of 31-51, which is significantly faster than existing NeRF-based works: Eikonal~\cite{bemana2022eikonal} (0.03 FPS) and NU-NeRF~\cite{sun2024nu} (0.016 FPS).

\begin{figure}[!t]
  \centering
  \includegraphics[width=0.98\linewidth]{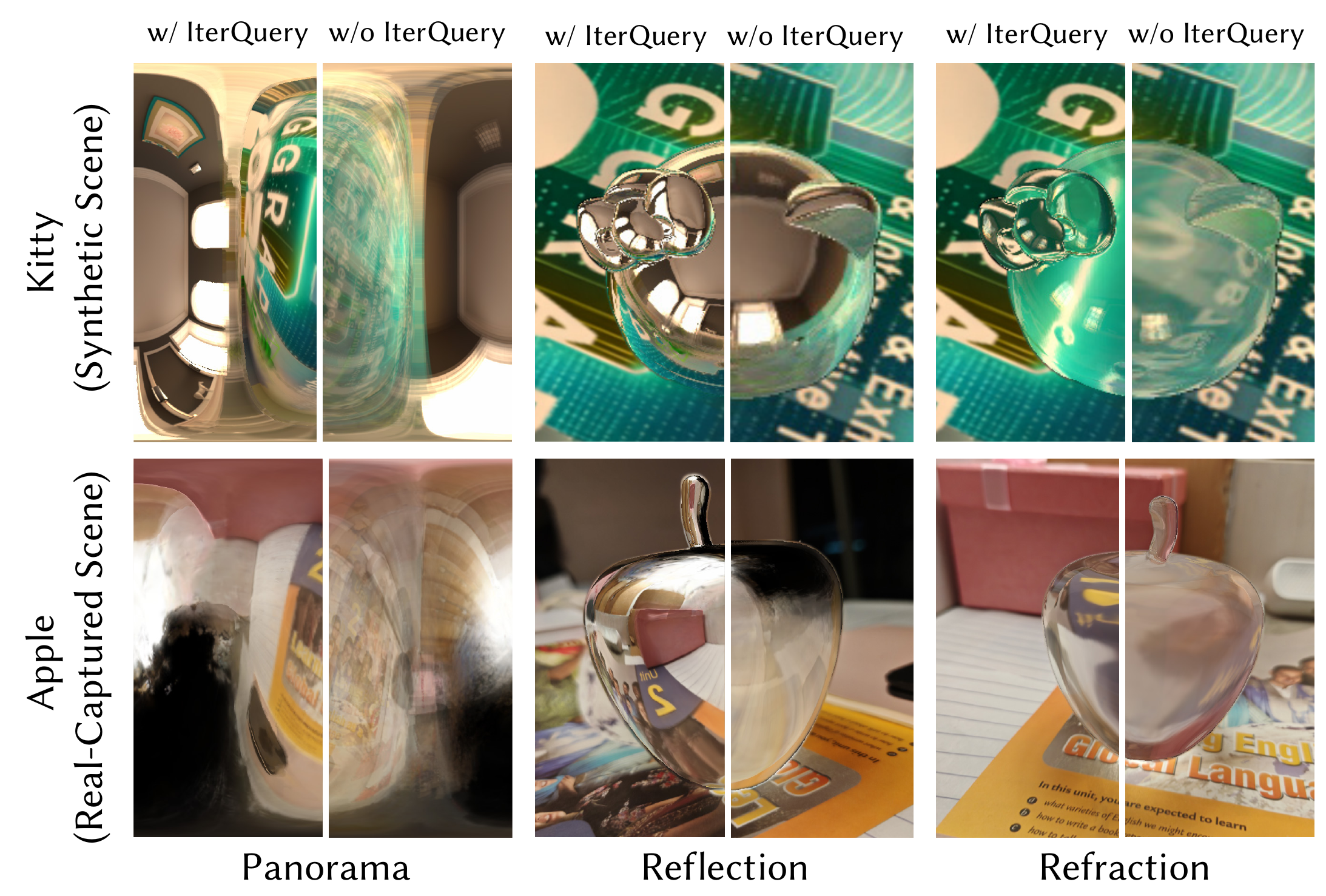} 
  \caption{\label{fig:exp_probes}
    \textbf{The impact of our IterQuery algorithm on rendering panoramas, reflection, and refraction.} Left half of each image: with IterQuery; Right half: without IterQuery. For the reflection, since the window is far enough from \textsc{Kitty}, the parallax effect is small, resulting in only slight visual differences.
  }
\end{figure}
\begin{figure*}[!t]
  \centering
  \includegraphics[width=1\linewidth]{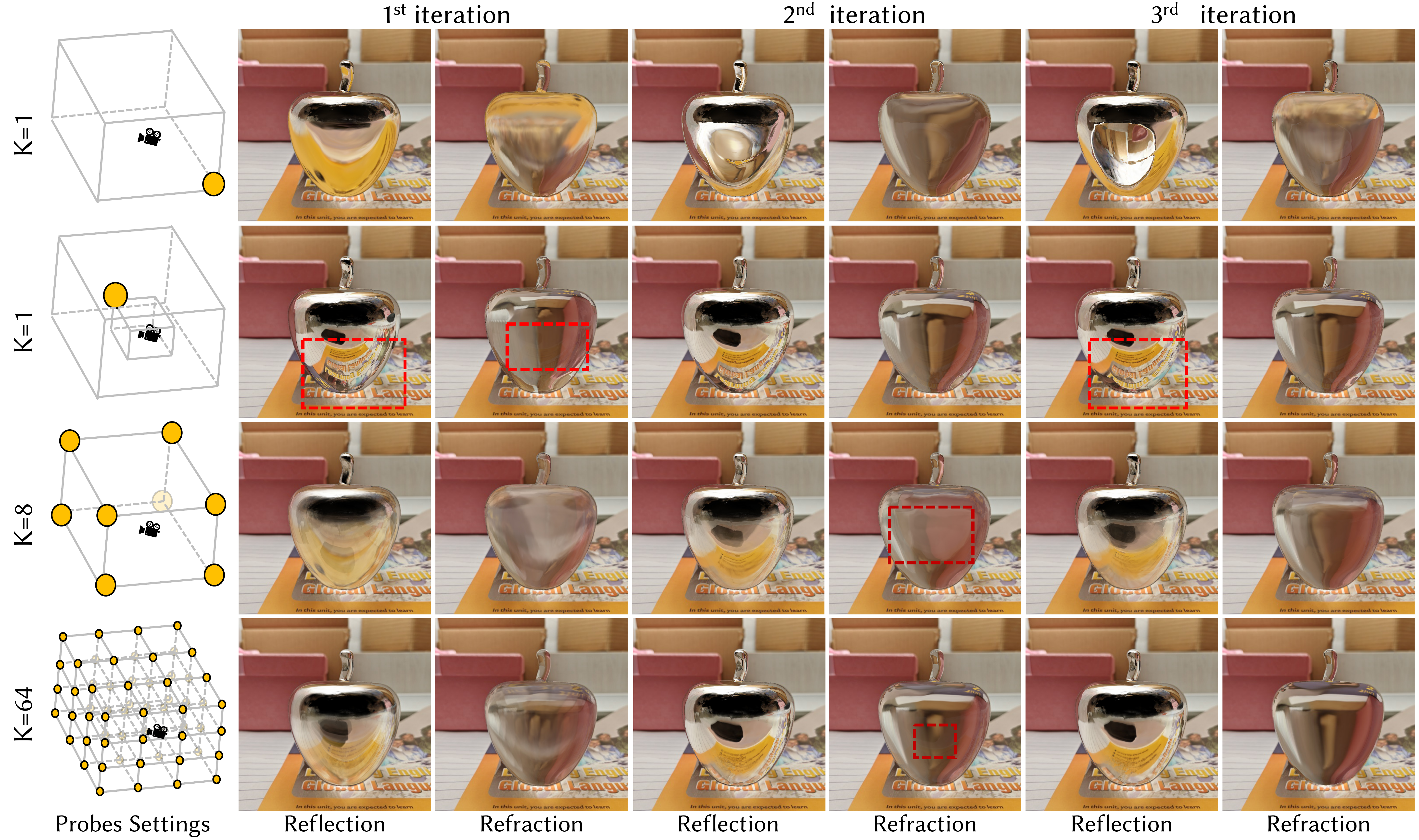} 
  \caption{\label{fig:discuss_probes}
  \textbf{
The impact of probes configurations on the convergence of the IterQuery algorithm.}  The leftmost image illustrates the number of the probes $K$ and the placement of the probes. In the 3rd iteration of the second row's reflection, it appears clearer but is incorrect, as only one line of text should be reflected at the bottom (red box).}
\end{figure*}

\subsection{Ablation Studies}

In this section, we conduct various ablation studies to validate the impact of key components in our method.

\paragraph{Ablation studies on inverse rendering} The key components of our method for inserse rendering of transparent objects are the deferred refraction strategy and our Gaussian light field probes. Thus, we provide quantitative and qualitative ablation studies under different settings on the \textsc{Kitty} scene, as reported in Tab.~\ref{tab:ablation} and shown in Fig.~\ref{fig:exp_defer}. It can be observed that the forward shading strategy blurs the high-frequency details of refraction and reflection in the rendering results. Without using probes, the overall color of the refracted light from the nearby table deviates significantly from the ground truth, but the high-frequency reflected light of ambient light from an “infinite distance” are preserved. Simultaneously employing deferred shading and Gaussian light field probes effectively combines the benefits of preserving high-frequency details and ensuring accurate query directions. The metrics further tell that our full method achieves the highest quality.

\paragraph{Ablation studies on the IterQuery algorithm} 

Given that the IterQuery algorithm plays a critical role in modeling indirect inter-reflection and refraction, we further conduct ablation studies on the algorithm. 

We visualize the rendered panoramas, reflection maps, and refraction maps using Gaussian light field probes in Fig.~\ref{fig:exp_probes}. It can be observed that without IterQuery, both the panoramas and refraction maps are over-blurred due to the parallax introduced by the probes. The reflected light from the window at the top of \textsc{Kitty} is sufficiently far away, making the parallax introduced by the probes negligible. However, the inter-reflected light at the bottom of \textsc{Kitty} is also blurred due to its proximity to the tabletop. On the contrary, our IterQuery algorithm effectively addresses the parallax issue and enhances the details of rendering. 

Different probes configurations, such as the count of iterations, the number and the placement of probes, also influence the convergence of the iteration and the rendering results, as illustrated in Fig.~\ref{fig:discuss_probes}. When \( K=1 \), the convergence is highly sensitive to the placement of the probe. The limited receptive field of the single probe makes it difficult to converge to the correct solution. As the iterations progress, the rendering results for the cases $K=8,64$ improve steadily, as indicated by the red boxes. Moreover, it can be observed that the case $K=64$ converges more rapidly. Theoretically, as \( K \) approaches infinity—where each incoherent ray is rasterized individually during the first iteration—the result converges to the optimal solution immediately. This provides a clear explanation for why larger $K$ values require fewer iterations for convergence. 

\begin{figure}[!t]
  \centering
  \includegraphics[width=0.95\linewidth]{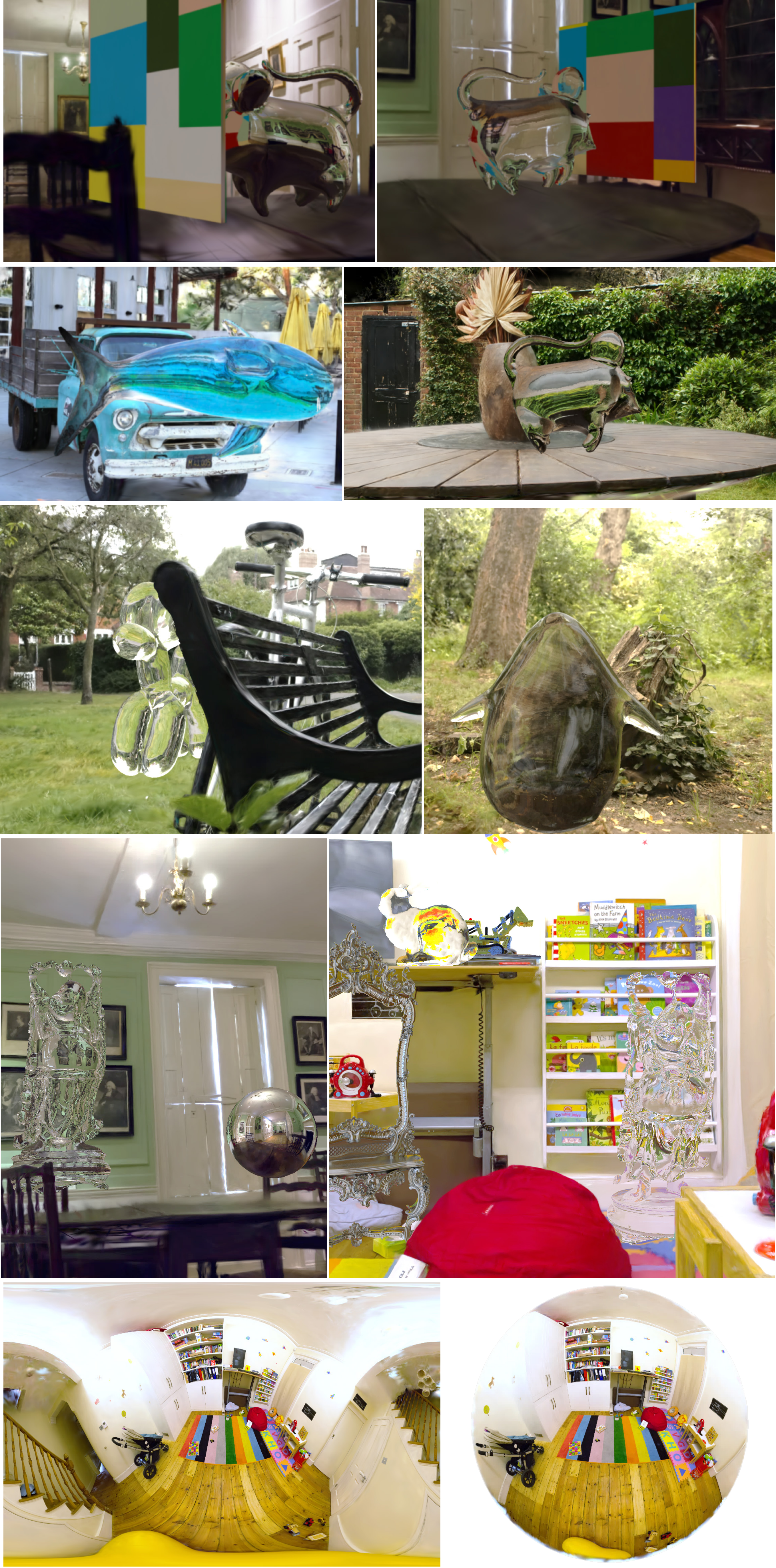} 
  \caption{\label{fig:exp_vt_fusion}
    \textbf{Applications.} Our method supports the rendering and navigation of scenes that integrate triangle meshes, traditional 3D-GS~\cite{kerbl20233d} and transparent Gaussian primitives, as well as non-pinhole cameras.
  }
\end{figure}

\subsection{Applications} 

Efficient Gaussian light fields open the door to many advanced applications, including secondary ray effects like mirrors, refractions, as well as rendering with non-pinhole cameras, as shown in Fig.~\ref{fig:exp_vt_fusion}. Please review our supplemental video for more details.

\paragraph{Re-rendering of reconstructed transparent objects} We place the transparent object \textsc{Mouse} reconstructed with our TransparentGS into the scene \textsc{DrJohnson}~\cite{hedman2018deep} reconstructed with the original 3D-GS~\cite{kerbl20233d} for re-rendering. Thanks to the reconstruction quality of TransparentGS and the query quality of GaussProbe, we can obtain photorealistic results for relighting and material editing. Additionally, our method demonstrates a significant performance advantage over other NeRF-based approaches. 

\paragraph{Secondary ray effects of inserted meshes} We use traditional triangle meshes (\textsc{Ball}, \textsc{StanfordBunny}, \textsc{Happy}, \textsc{Mirror}, and \textsc{Luck}, the colored plane), 3D-GS~\cite{kerbl20233d} (\textsc{DrJohnson}, \textsc{Playroom}~\cite{hedman2018deep}, \textsc{Truck}~\cite{knapitsch2017tanks}, 
\textsc{Stump},
\textsc{Bicycle},
\textsc{Garden}~\cite{barron2022mip}, and \textsc{Lego}~\cite{mildenhall2020nerf}), and transparent Gaussian primitives (\textsc{Mouse}, \textsc{Apple}, \textsc{Penguin}, \textsc{Dog}, and \textsc{Dolphin}) to create a variety of scenes. GaussProbe enable secondary ray effects in these multi-object composition scenes, as shown in Fig.~\ref{fig:teaser} and Fig.~\ref{fig:exp_vt_fusion}, which is challenging for other methods.

\paragraph{Rendering with non-pinhole camera models} By the way, we can also leverage GaussProbe for rendering with non-pinhole cameras, such as \textsc{Playroom}~\cite{hedman2018deep} with fisheye cameras and panoramas.

\subsection{Discussions and Limitations}

\begin{figure}[!t]
  \centering
  \includegraphics[width=0.92\linewidth]{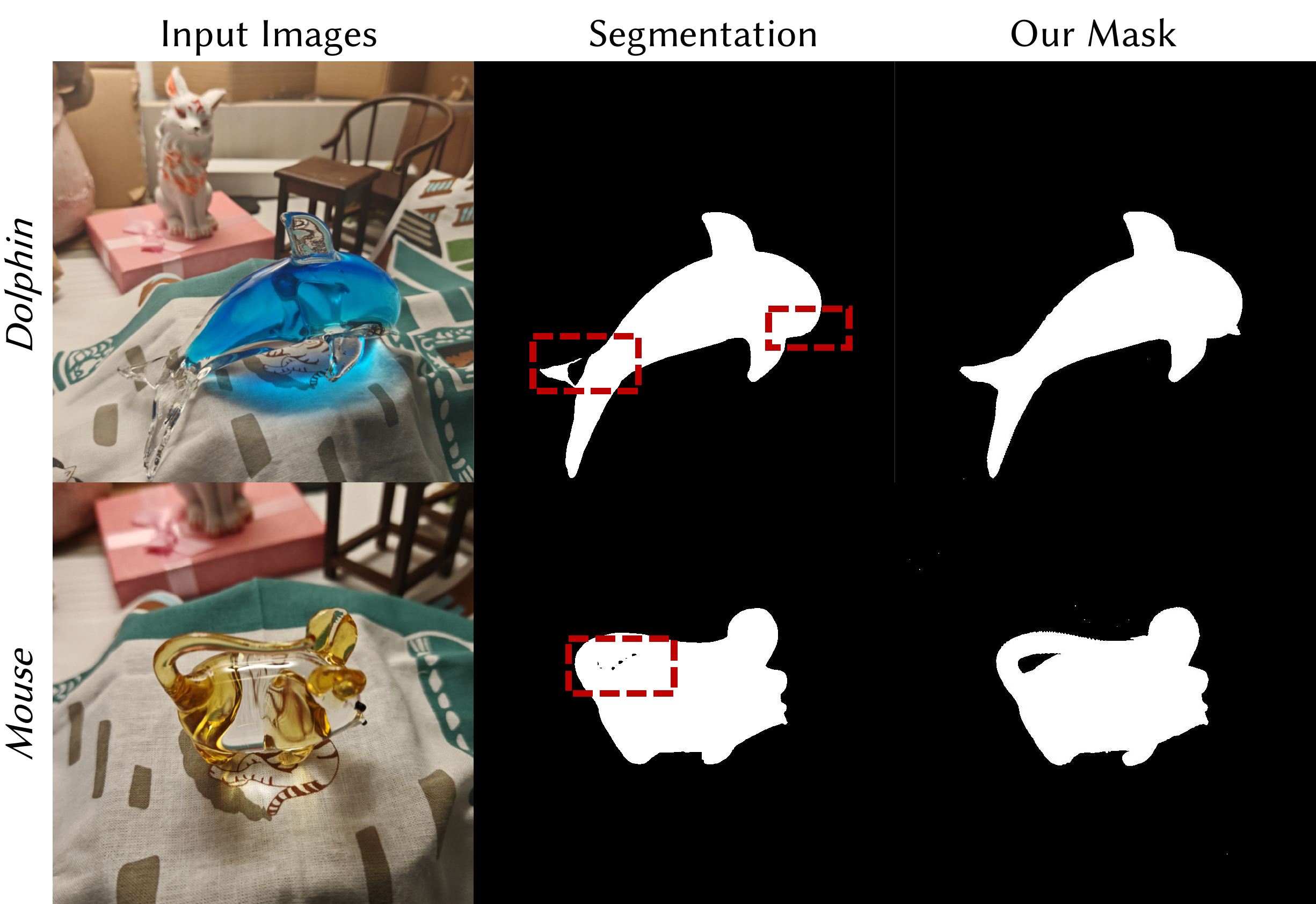} 
  \caption{\label{fig:robust_sam}
    \textbf{Robustness of our method to segmentation results. }  The red boxes highlight the regions with imprecise segmentation results.
  }
\end{figure}

\paragraph{Robustness of our method to segmentation results} Since we use semantic segmentation to partition the scene, we validate the impact of segmentation results on the reconstruction of transparent objects, as shown in Fig.~\ref{fig:robust_sam}. It can be observed that although the segmentation results are mostly accurate, there are still imprecise regions, such as the dolphin's mouth and the mouse's tail. However, our method demonstrates sufficient robustness to such inaccuracy through supervision from multiple views.

\begin{figure}[!t]
  \centering
  \includegraphics[width=0.95\linewidth]{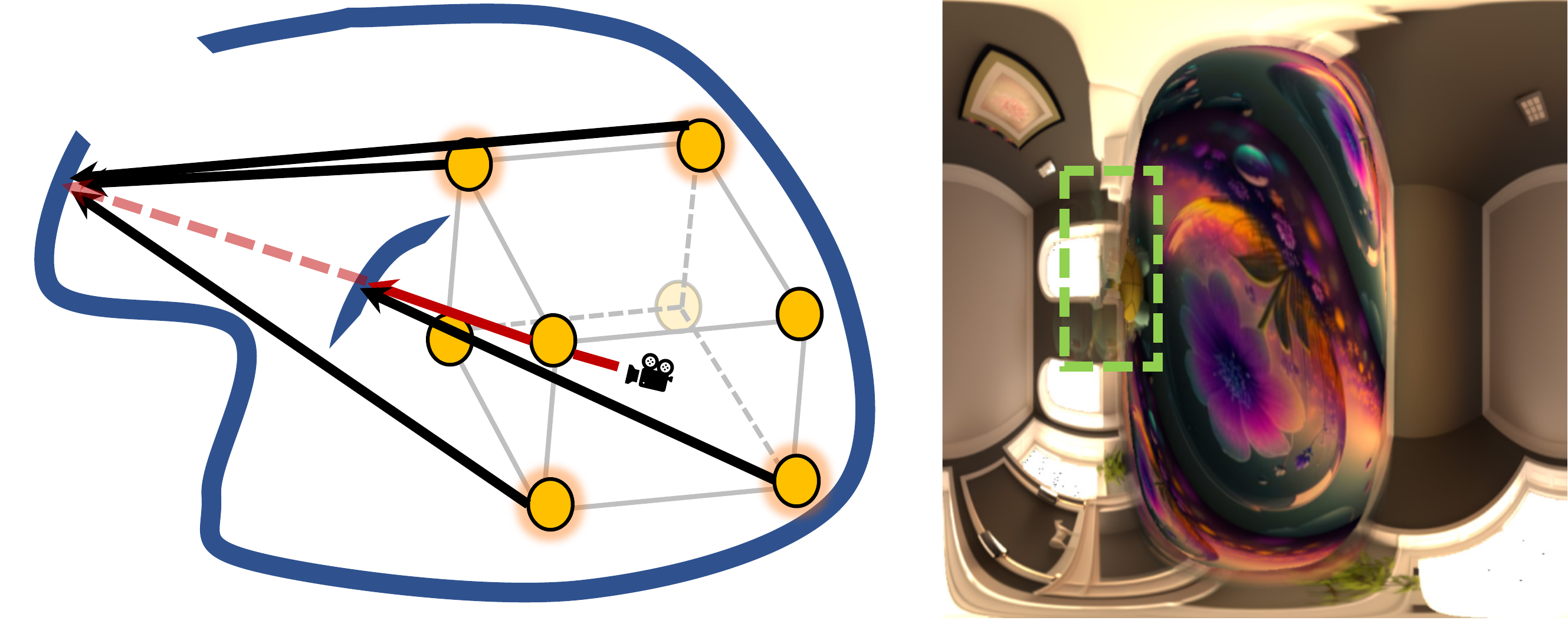} 
  \caption{\label{fig:limitation}
    \textbf{Failure cases for the IterQuery algorithm.} The green box highlights the regions where our IterQuery algorithm fails when there are non-negligible singularities that clearly violate the manifold constraint.
  }
\end{figure}

\paragraph{Manifold constraint} For our Gaussian light field probes, a single probe with the depth panorama can represent a two-dimensional manifold embedded in 3D space. The IterQuery algorithm can be viewed as the process of a moving point sliding along the manifold until it coincides with the query ray. However, as illustrated in Fig.~\ref{fig:limitation}, there are cases where noticeable discontinuities occur, violating the manifold constraint. In such cases, it is possible that not all probes converge to the first intersection between the rays to be queried and the 3D scene, leading to incorrect results. This also offers an explanation for the failure case $K=1$, as illustrated in Fig.~\ref{fig:discuss_probes}. Fortunately, this can be resolved by increasing the number of probes and employing a more optimal placement strategy.

\begin{figure}[!t]
  \centering
  \includegraphics[width=0.92\linewidth]{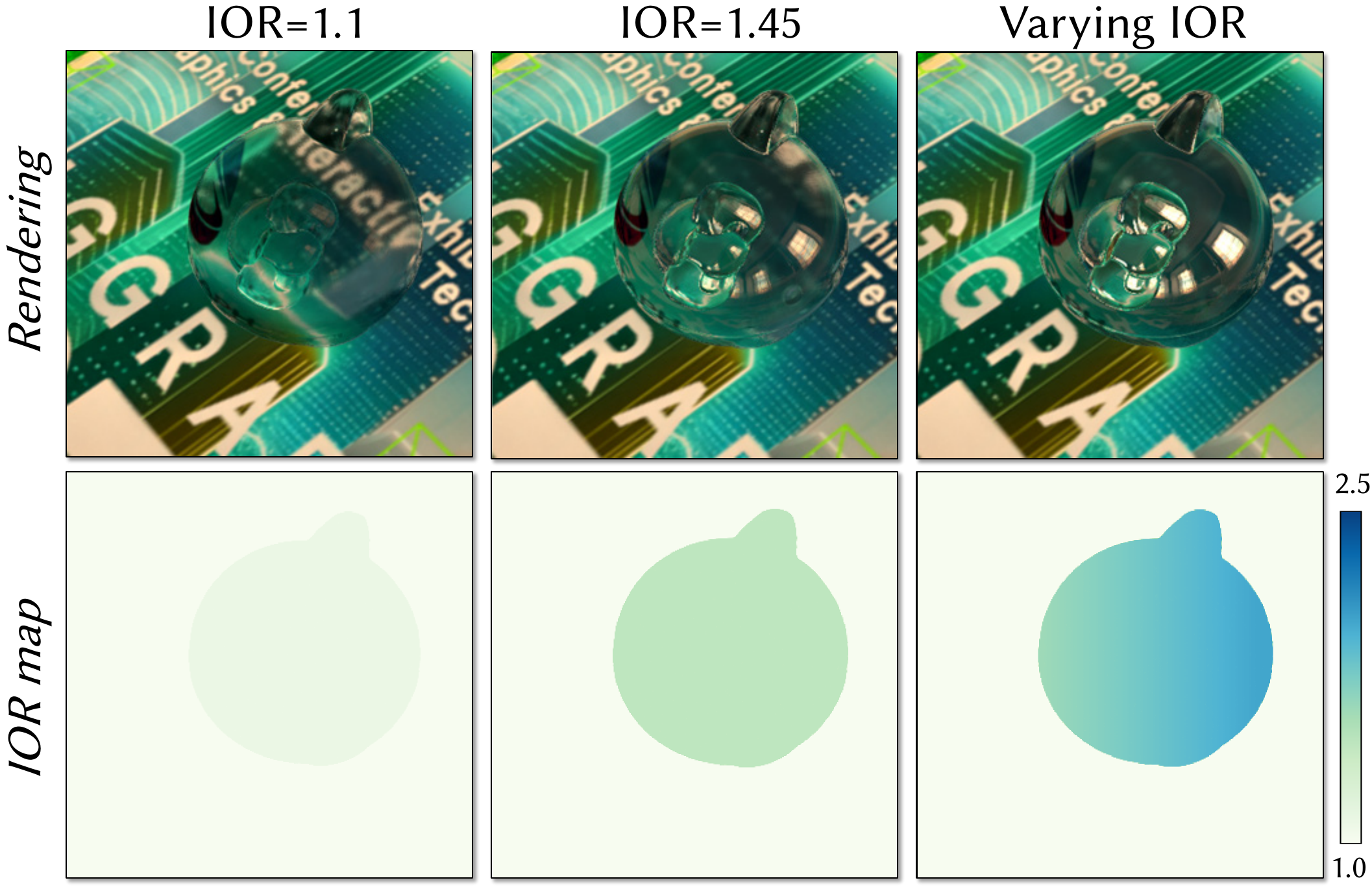} 
  \caption{\label{fig:varyior} 
  \textbf{Transparent objects with different IORs.} Higher index of refraction values are depicted in blue, and lower values are depicted in green.}
\end{figure}

\paragraph{Complex light paths and ambiguity} The light paths are crucial for simulating how light behaves in a transparent object. Similar to previous methods~\cite{gao2023transparent, li2023neto, li2020through}, we restrict the number of light bounces. Specifically, we assume that a light path consists of exactly two refractions, with at most one total internal reflection, such as \textsc{HalfBall} in Fig.~\ref{fig:comp}. However, complex light paths with more bounces can introduce ambiguity and singularity, posing challenges for inverse rendering. Therefore, accurately reconstructing complex geometries, such as hollow transparent objects and those with intricate self-occlusion, remains challenging when using consumer-level cameras. Furthermore, although our method supports varying IOR (Fig.~\ref{fig:varyior}), highly heterogeneous transparent objects may introduce ambiguity in material estimation. This may be
addressed by a generative model, which we leave as future work. 

\paragraph{Invisible environment}

Similar to existing inverse rendering methods based on NeRF~\cite{mildenhall2020nerf} or 3D-GS~\cite{kerbl20233d} (e.g., Eikonal~\cite{bemana2022eikonal} or GShader~\cite{jiang2024gaussianshader}), incomplete environment will affect the reconstruction of invisible parts. Therefore, to improve the accuracy of reconstruction, it is better to capture the whole environment.

\paragraph{Gaussian ray tracing}

We didn't choose 3D Gaussian Ray Tracing (3DGRT)~\cite{moenne20243dtracing} in our pipeline due to the following reasons. Firstly, 3DGRT currently does not support inverse rendering and material decomposition, as claimed in its limitations. Secondly, unlike 3DGRT, our GaussProbe does not need to build a BVH or maintain a sorted buffer, making it more efficient for our task. Moreover, GaussProbe can be directly obtained from scenes trained with the original 3DGS, whereas 3DGRT requires fine-tuning. That said, it would be an interesting future work to extend 3DGRT to reconstruct transparent objects.

\paragraph{Caustics rendering}

Since our task is inverse rendering, we focus more on the influence of the surrounding on the object's appearance than the influence of the object on its surrounding (e.g., caustics). Currently, 3DGS-based methods fail to handle caustics rendering. It would be an interesting future work.

\paragraph{Video}

We provide a video in the supplementary material. Some visual artifacts in the video primarily arise from the aliasing caused by sampling, the accuracy of the marching cube algorithm, and the moving object bounding boxes. These issues can potentially be mitigated by increasing the samples per pixel, and increasing the coverage of the probes to fully encompass the area within the dynamic object's movement range, etc. 


\section{Conclusion}

In conclusion, we propose TransparentGS, an innovative and efficient inverse rendering framework for transparent objects based on 3D-GS. By introducing the transparent Gaussian primitives, which incorporate material attributes alongside positions and shapes, our method enables physically-based rendering of transparent materials. To preserve refractive details, we leverage a deferred refraction strategy, to effectively capture specular refraction. Addressing the challenges posed by the high computational cost of ray tracing and the limitations of rasterization in handling incoherent secondary rays, we employ the Gaussian light field probes (GaussProbe) to represent nearby and distant light variations, accompanied by an IterQuery algorithm that mitigates the parallax artifacts in probe query. The GaussProbe-based pipeline allows us to reconstruct any transparent object within one hour and facilitates realistic rendering. It opens the door to complex secondary ray effects in interplay of the multi-object composition scene. Experimental results confirm the effectiveness and efficiency of our framework, highlighting its potential for advancing transparent objects.

\begin{acks}
We would like to thank the anonymous reviewers for their valuable feedback. We also thank the authors of \citet{zhou2024unified} for generously providing the LaTeX template of the pseudocode. This work was supported by the National Natural Science Foundation of China (No. 61972194 and No. 62032011) and the Natural
Science Foundation of Jiangsu Province (No. BK20211147).
\end{acks}

\bibliographystyle{ACM-Reference-Format}
\bibliography{main}

\end{document}